\documentclass[UTF8,a4paper,twocolumn]{aastex61}

\usepackage{amsmath}

\newcommand{\oh}{12\,+\,log(O/H)}

\newcommand{\vs}{vs.}

\newcommand{\ha}{\hbox{H$\alpha$}}

\newcommand{\hi}{\hbox{H\,{\sc i}}}
\newcommand{\hii}{\hbox{H\,{\sc ii}}}

\newcommand{\gsim}{\lower.5ex\hbox{$\; \buildrel > \over \sim \;$}}
\newcommand{\lsim}{\lower.5ex\hbox{$\; \buildrel < \over \sim \;$}}

\newcommand{\um}{$\mu$m}

\submitjournal{AAS}

\shorttitle{PAH Emission as a Function of Age}
\shortauthors{Lin et al.}

\begin{document}

\title{The Age-Dependence of Mid-Infrared Emission Around Young Star Clusters}

\email{zesenlin@mail.ustc.edu.cn, calzetti@astro.umass.edu}

\author[0000-0001-8078-3428]{Zesen Lin}
\affil{Key Laboratory for Research in Galaxies and Cosmology, Department of Astronomy, University of Science and Technology of China, Hefei 230026, China}
\affil{School of Astronomy and Space Sciences, University of Science and Technology of China, Hefei, 230026, China}
\affil{Department of Astronomy, University of Massachusetts, Amherst, MA 01003, USA}

\author[0000-0002-5189-8004]{Daniela Calzetti}
\affil{Department of Astronomy, University of Massachusetts, Amherst, MA 01003, USA}

\author[0000-0002-7660-2273]{Xu Kong}
\affil{Key Laboratory for Research in Galaxies and Cosmology, Department of Astronomy, University of Science and Technology of China, Hefei 230026, China}
\affil{School of Astronomy and Space Sciences, University of Science and Technology of China, Hefei, 230026, China}

\author[0000-0002-8192-8091]{A. Adamo}
\affil{Department of Astronomy, Oskar Klein Centre, Stockholm University, AlbaNova University Centre, SE-106 91 Stockholm, Sweden}

\author[0000-0001-6291-6813]{M. Cignoni}
\affil{Department of Physics, University of Pisa, Largo Pontecorvo, 3, I-56127 Pisa, Italy}

\author[0000-0002-6877-7655]{D. O. Cook}
\affil{IPAC/Caltech, Pasadena, CA 91101, USA}

\author[0000-0002-5782-9093]{D. A. Dale}
\affil{Department of Physics and Astronomy, University of Wyoming, Laramie, WY, USA}

\author[0000-0002-3247-5321]{K. Grasha}
\affil{Australian National University, Canberra, Australia}

\author[0000-0002-1891-3794]{E. K. Grebel}
\affil{University of Heidelberg, Germany}

\author[0000-0003-1427-2456]{M. Messa}
\affil{Department of Astronomy, University of Massachusetts, Amherst, MA 01003, USA}

\author[0000-0001-5618-0109]{E. Sacchi}
\affil{Space Telescope Science Institute, 3700 San Martin Drive, Baltimore, MD 21218, USA}

\author[0000-0002-0806-168X]{L. J. Smith}
\affil{European Space Agency/Space Telescope Science Institute, Baltimore, MD, USA}




\begin{abstract}

    Using the star cluster catalogs from the Hubble Space Telescope program Legacy ExtraGalactic UV survey (LEGUS) and 8 \um\ images from the IRAC camera on the Spitzer Space Telescope for five galaxies within 5 Mpc, we investigate how the 8 \um\ dust luminosity correlates with the stellar age on the 30--50 pc scale of star forming regions. We construct a sample of 97 regions centered at local peaks of 8 \um\ emission, each containing one or more young star cluster candidates from the LEGUS catalogs. We find a tight anticorrelation with a Pearson correlation coefficient of $r=-0.84\pm0.05$ between the mass-normalized dust-only 8 \um\ luminosity and the age of stellar clusters younger than 1 Gyr; the 8 \um\ luminosity decreases with increasing age of the stellar population. Simple assumptions on a combination of stellar and dust emission models reproduce the observed trend. We also explore how the scatter of the observed trend depends on assumptions of stellar metallicity, polycyclic aromatic hydrocarbon (PAH) abundance, fraction of stellar light absorbed by dust, and instantaneous versus continuous star formation models. We find that variations in stellar metallicity have little effect on the scatter, while PAH abundance and the fraction of dust-absorbed light bracket the full range of the data. We also find that the trend is better explained by continuous star formation, rather than instantaneous burst models. We ascribe this result to the presence of multiple star clusters with different ages in many of the regions. Upper limits of the dust-only 8 \um\ emission as a function of age are provided.

\end{abstract}




\section{Introduction}
\label{sec:intro}

    The star formation rate (SFR) is one of the key parameters that trace the evolution of galaxies through cosmic time. A number of studies have published calibrations of SFR indicators that cover a wide wavelength range from ultraviolet (UV) to radio (e.g., \citealt{Kennicutt1998,Kennicutt2012}).

    Monochromatic infrared (IR) indicators, mainly those based on the bands of the Spitzer Space Telescope or the Herschel Space Observatory, have been calibrated as tools to capture the dust-obscured SFR, while avoiding the observational complexity of observing the bolometric infrared emission (e.g., \citealt{Wu2005,Calzetti2007,Calzetti2010,Brown2017}). Among these indicators, the rest-frame 8 \um\ luminosity is of particular interest for studies of high-redshift galaxies (e.g., \citealt{Reddy2006,Reddy2012,Elbaz2011,Shivaei2017}) for two reasons: (1) the 8 \um\ emission region is dominated by strong emission features from polycyclic aromatic hydrocarbons (PAHs; \citealt{Leger1984,Allamandola1985}), which are easily detectable at large distances; and (2) the rest-frame 8 \um\ wavelength region of galaxies is observable out to redshift $z\sim2-2.5$ with mid-IR capabilities, such as the recent-past MIPS 24 \um\ on Spitzer \citep{Rieke2004} and the upcoming Mid-Infrared Instrument (MIRI) on the James Webb Space Telescope (JWST; \citealt{Bouchet2015,Rieke2015}). The correlation between the 8 \um\ luminosity and SFR is supported by studies at multiple wavelengths, including \ha\ \citep{Wu2005,Kennicutt2009}, Paschen-$\alpha$ \citep{Calzetti2005,Calzetti2007}, the bolometric IR \citep{Reddy2006,Shivaei2017}, and even radio wavelengths (e.g., \citealt{Wu2005}). However, the 8 \um\ luminosity is also not directly proportional to the amount of ionizing emission \citep{Calzetti2005}, due to the fact that a large fraction of 8 \um\ emission is excited by nonionizing stellar populations, which include the field populations of galaxies \citep{Calzetti2007,Draine2007,Smith2007,Crocker2013,Lu2014}.

    The dust emission around the rest-frame 8 \um\ comprise the underlying continuum excited by hot, stochastically heated small dust grains and emission features generally attributed to the PAHs \citep{Draine2007,Smith2007,Tielens2008}. In star-forming regions, the PAHs mainly reside in the photodissociation region (PDR) where the atomic-to-molecular (e.g., \hi-to-H$_2$) phase transition takes place \citep{Tielens2013,Bialy2016}. Observations of luminous \hii\ regions in M33 revealed that the 8 \um\ emission is located between \ha\ (ionized gas) and CO emission (molecular gas), indicating the location of PDRs \citep{Relano2009}. For dust heated by star formation activity, the 8 \um\ luminosity is dominated by emission from PAHs \citep{Povich2007,Smith2007,Marble2010,Tielens2013}. As such, the dust emission within the Spitzer IRAC4 (8 \um) band will be approximately attributed to the PAHs in this work for simplicity.

    Observations of nearby star-forming regions/galaxies revealed that there is a metallicity dependence for the PAH emission \citep{Engelbracht2005,Madden2006,Smith2007,Gordon2008} or abundance \citep{Draine2007a,Remy-Ruyer2015}, with weaker (lower) PAH emission (abundance) for more metal-poor regions/galaxies. Conversely, the existence of a strong interstellar radiation field is found to suppress the PAH emission, regardless of the metallicity \citep{Madden2006,Gordon2008,Lebouteiller2011,Shivaei2017,Binder2018}. Such correlations between the PAH emission strength and the environment were explained in terms of PAH formation (e.g., \citealt{Sandstrom2012}) or destruction (e.g., \citealt{Madden2006,Gordon2008,Lebouteiller2011,Binder2018}). \cite{Galliano2008} suggested that the delayed injection of carbon dust by asymptotic giant branch (AGB) stars can explain the observed metallicity-dependent PAH abundance. This picture suggests a positive relation between the stellar age and the PAH abundance that was supported by \cite{Shivaei2017} for star-forming galaxies (SFGs) at $z\sim2$ in which galaxies with $\mathrm{ages}\gtrsim 900$ Myr exhibit an elevated PAH-to-IR luminosity ratio compared galaxies with younger populations.

    While the environment regulates the PAH abundance, the heating sources significantly alter the PAH emission at fixed PAH abundance. Spatially resolved studies demonstrated that a large amount of 8 \um\ emission is associated with the diffuse cold interstellar medium \citep{Bendo2008,Calapa2014}, suggesting a heating source other than recent star formation. The reported fractions of flux in the IRAC4 band that are not related to recent star formation and are excited by evolved stars are about 30\%--40\%, 67\%, and 60\%--80\% in NGC 628 \citep{Crocker2013}, M81 \citep{Lu2014}, and M33 \citep{Calapa2014}, respectively. These observations suggest that one should address how the PAH emission behaves as a function of the age of the stellar population(s) that heat/excite them before applying the 8 \um\ emission as an SFR tracer.

    In this work, we aim to investigate how the 8 \um\ emission correlates with the age of the stellar population that heats the dust locally utilizing the star cluster catalogs from the Legacy ExtraGalactic UV survey (LEGUS; \citealt{Calzetti2015a}). LEGUS is a treasury program of the Hubble Space Telescope (HST) aimed at the investigation of star formation and its relation with the galactic environment in 50 nearby galaxies within the local 16 Mpc. The LEGUS observations cover five broad bands from the UV to the near-IR (NIR) using the Wide Field Camera 3 or the Advanced Camera for Surveys on board the HST. Based on these high spatial resolution images, $\sim10,000$ young star clusters (YSCs) were extracted and measured in 37 pointings in 31 galaxies \citep{Adamo2017,Cook2019}.

    The paper is organized as follows. We describe the YSC selections for this analysis in Section \ref{sec:sample}, and the image processing and photometry for IRAC4 sources in Section \ref{sec:image}. In Section \ref{sec:results} we present the observed relation between the star cluster age and the mass-normalized 8 \um\ luminosity. Simple assumptions are used to combine publicly available models of stellar populations and dust emission in order to understand the observed trend and the scatter in Section \ref{sec:models}. In Section \ref{sec:discussion} we discuss reasons for the observed deviation of several sources from the general trend and present an analytic expression for the envelope of the maximum 8 \um\ emission at a given age. Finally, we summarize our work in Section \ref{sec:summary}. Throughout this paper, we adopt a \cite{Kroupa2001} initial mass function (IMF).

\section{Sample Selection}
\label{sec:sample}

    \subsection{Galaxy Selection}
    \label{ssec:galaxy_selection}

\begin{figure*}[htb]
        \centering
        \includegraphics[width=\textwidth]{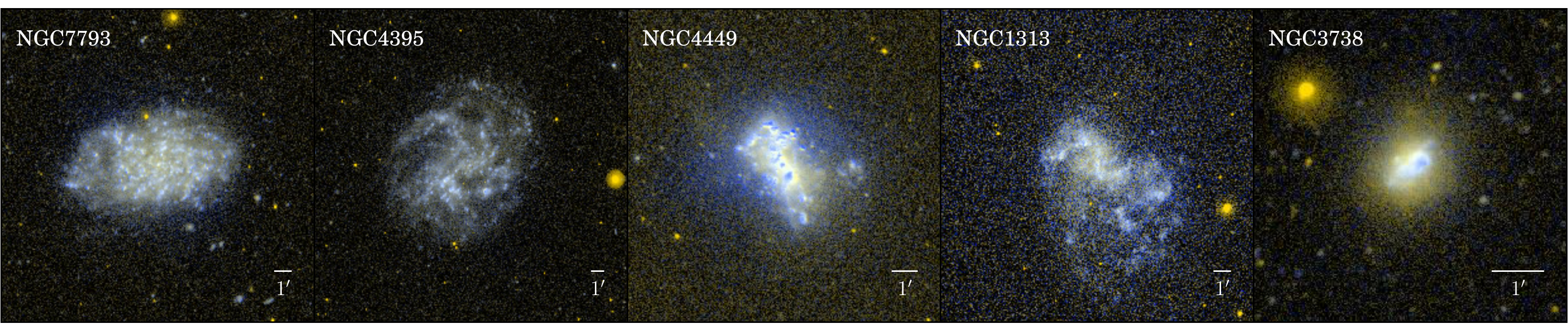}
        \caption{GALEX two-color (far-UV and near-UV) images for our five galaxies. The name of each galaxy is listed at the upper-left corner of each panel. The white bar indicates the 1\arcmin\ scale for each panel.
        \label{fig:galex_image}}
        \end{figure*}

        The LEGUS sample consists of 50 galaxies between 3.5 and $\sim16$ Mpc distance. Within this distance range, the Spitzer/IRAC camera, with a $\sim2\arcsec$ point spread function (PSF) subtends regions between 30 and 150~pc. In order to maximize the number of isolated star clusters within each IRAC PSF, we choose a maximum distance of 5 Mpc for the most distant galaxy we analyze, which subtends regions $\sim50$ pc in size or the typical size of \hii\ regions. In addition, we require the LEGUS catalogs to include at least 100 identified star cluster candidates (class 1, 2, or 3; see below) to ensure sufficient statistics. This reduces our sample to the 5 galaxies listed in Table \ref{tab:properties}. In this table we list, in addition to cluster numbers, also several basic physical properties for each galaxy, such as morphologies, adopted distances, oxygen abundances, and stellar masses. In Figure \ref{fig:galex_image} we display the Galaxy Evolution Explorer (GALEX) 2-color (far-UV and near-UV) images for these galaxies, which are obtained from the GALEX GR6/7 Data Release\footnote{http://galex.stsci.edu/gr6/}.

        \begin{deluxetable*}{cccccc}[htb]
        \tablecaption{Properties of Selected Galaxies\label{tab:properties}}
        \tablecolumns{6}
        \tablewidth{0pt}
        \tablehead{
        \colhead{Properties} & \colhead{NGC 7793} & \colhead{NGC 4395} & \colhead{NGC 4449} & \colhead{NGC 1313} & \colhead{NGC 3738}}
        \startdata
        Morph.\tablenotemark{a} & SAd & SAm & IBm & SBd & Im \\
        Dist. (Mpc)\tablenotemark{b} & 3.44 & 4.30 & 4.31 & 4.39 & 4.90 \\
        \oh\tablenotemark{c} & 8.50 & 8.19 & 8.26 & 8.21 & 8.10 \\
        $M_*$ ($M_{\odot}$)\tablenotemark{d} & $3.2\times10^9$ & $6.0\times10^8$ & $1.1\times10^9$ & $2.6\times10^9$ & $2.4\times10^8$ \\
        $N$(YSC)\tablenotemark{e} & 378 & 108 & 284 & 674 & 164 \\
        $N$(IRAC4 clumps)\tablenotemark{f} & 48 & 17 & 25 & 47 & 9 \\
        $N$(final)\tablenotemark{g} & 39 & 17 & 16 & 23 & 2 \\
        \enddata
        \tablenotetext{a}{Morphological type as listed in the NASA/IPAC Extragalactic Database.}
        \tablenotetext{b}{Redshift-independent or flow-corrected redshift-dependent distance in Mpc, taken from \cite{Calzetti2015a}.}
        \tablenotetext{c}{Central oxygen abundances of the galaxies derived from the empirical calibrations \citep{Pilyugin2005,Pilyugin2012,Pilyugin2013} or the direct method \citep{Pilyugin2010,Pilyugin2012}. The values for NGC 7793, NGC 4395, and NGC 1313 are taken from \cite{Pilyugin2014}, while the ones for NGC 4449 and NGC 3738 are taken from \cite{Pilyugin2015}.}
        \tablenotetext{d}{Stellar masses in $M_{\odot}$, taken from \cite{Calzetti2015a}.}
        \tablenotetext{e}{Number of identified YSCs; see Section \ref{ssec:ysc_catalogs}.}
        \tablenotetext{f}{Number of IRAC4 sources identified as YSC counterparts; see Section \ref{ssec:irac4_photometry}.}
        \tablenotetext{g}{Number of IRAC4 sources in the final sample; see Section \ref{sec:results}.}
        \end{deluxetable*}

    \subsection{YSC Catalogs}
    \label{ssec:ysc_catalogs}

        We take the LEGUS catalogs from the public repository at \url{https://archive.stsci.edu/prepds/legus/dataproducts-public.html} \citep{Adamo2017,Cook2019} as the parent sample, which include photometry in five bands (NUV, $U$, $B$, $V$, $I$) and physical properties such as stellar mass ($M_*$), age, and extinction derived from fitting the spectral energy distribution (SED) of each cluster candidate \citep[][their Figure~9]{Calzetti2015a}. The physical parameters are derived using a traditional $\chi^2$ minimization approach between models and photometric data, as described in \cite{Adamo2010}. The more accurate Bayesian approach, in which model parameters are stochastically sampled from cluster evolutionary models, has been applied to the catalogs of only two LEGUS galaxies \citep{Krumholz2015}. For this reason, we use the physical parameters derived from the $\chi^2$ approach, also called the deterministic approach \citep{Adamo2017}.

        The SED fitting for the deterministic catalogs uses two stellar libraries---Padova-AGB \citep{Bressan1993,Fagotto1994,Fagotto1994a,Girardi2000,Vazquez2005} and Geneva tracks without rotation \citep{Ekstroem2012,Georgy2013,Leitherer2014}---and three extinction/attenuation laws using photometry from two types of aperture corrections (average based and concentration index (CI) based). The combination of these choices produces 12 catalogs for each LEGUS galaxy \citep{Adamo2017}. Here, we make use of the catalogs constructed from the average aperture corrections and SED fitting utilizing the Geneva tracks and Milky Way extinction law of \cite{Cardelli1989}. The choice of aperture correction or model assumption does not have a significant influence on our final conclusions, as briefly discussed in Section \ref{sec:discussion}. Extensive details about the fitting approach and parameter assumptions can be found in \cite{Adamo2017}.

        The LEGUS catalogs include visually inspected star cluster candidates assigned to ``class 1'' (symmetric, compact morphology), ``class 2'' (compact, but elongated), and ``class 3'' (multipeak). These have been inspected by at least three individuals, after ensuring that each cluster candidate is detected in at least four of the five LEGUS bands with photometric error $\sigma\leq0.3$ mag. The \cite{Adamo2017} catalogs use a $Q$ value, which is the probability that $\chi^2$ exceeds a particular value by chance \citep{Press2007}, to evaluate the quality of the fit; this parameter is larger than 0.1 if the fit is good \citep{Press2007,Adamo2010}. Note that the $Q$ parameter is not the reduced $\chi^2$ ($\chi_{\mathrm{red}}^2$), which is close to 1 for a good fit. To restrict the goodness of fit to the higher quality results, as suggested by \cite{Adamo2010}, we only include YSCs with $Q>0.001$ ($\sim80\%$ of cluster candidates with class=1, 2, or 3) in our sample, indicating that the best-fit models are acceptable \citep{Press2007}. The resulting sample has a median and 68\% scatter of $0.31_{-0.28}^{+0.49}$ and $1.20_{-0.87}^{+1.89}$ for $Q$ and $\chi_{\mathrm{red}}^2$, respectively. See \cite{Adamo2017} for more details.

\section{Image Processing and photometry}
\label{sec:image}

    \subsection{IRAC Image Processing}
    \label{ssec:irac_processing}

        The IRAC1 (3.6 \um) and IRAC4 (8.0 \um) images for our sample galaxies are taken from the Spitzer Local Volume Legacy (LVL; \citealt{Dale2009}) survey,\footnote{\url{https://irsa.ipac.caltech.edu/data/SPITZER/LVL/}} which provides IRAC and MIPS images for 258 galaxies out to 11 Mpc. Most of the LEGUS galaxies with public YSC catalogs are also included in the LVL sample by design. The IRAC images are resampled to a pixel scale of 0\farcs75, and the FWHMs of their PSFs are $\sim 1\farcs6$ and $\sim 1\farcs9$ for IRAC1 and IRAC4, respectively \citep{Dale2009}.

        The IRAC1 images are convolved to match the PSF of IRAC4 images using the Photutils package \citep{bradley2019}. Under the assumption that all the 3.6 \um\ emission is from stars, we remove stellar emission from the IRAC4 images using the formula of \cite{Helou2004}. A small fraction (about 5\%--33\%) of the 3.6 \um\ flux is due to 3.3 \um\ PAH emission \citep{Meidt2012,Robitaille2012a}. We ignore this small contribution, which is expected to have negligible impact ($\lesssim1.5\%$ for our sample) on the dust-only 8 \um\ images. 

    \subsection{IRAC4 Photometry}
    \label{ssec:irac4_photometry}

        The IRAC4 photometry is performed on the dust-only (star-subtracted) images utilizing the PHOT task in the PyRAF\footnote{PyRAF is a product of the Space Telescope Science Institute, which is operated by AURA for NASA.} interface of IRAF \citep{Tody1986,Tody1993}. Due to the large PSF of the IRAC4 images, most of the sources are blended with their neighbors. To extract accurate fluxes for each source, we limit our analysis to relatively isolated sources and perform photometry in small apertures with aperture corrections determined in a way similar to that described in \cite{Adamo2017}.

        \begin{figure*}[htb]
        \centering
        \includegraphics[width=\textwidth]{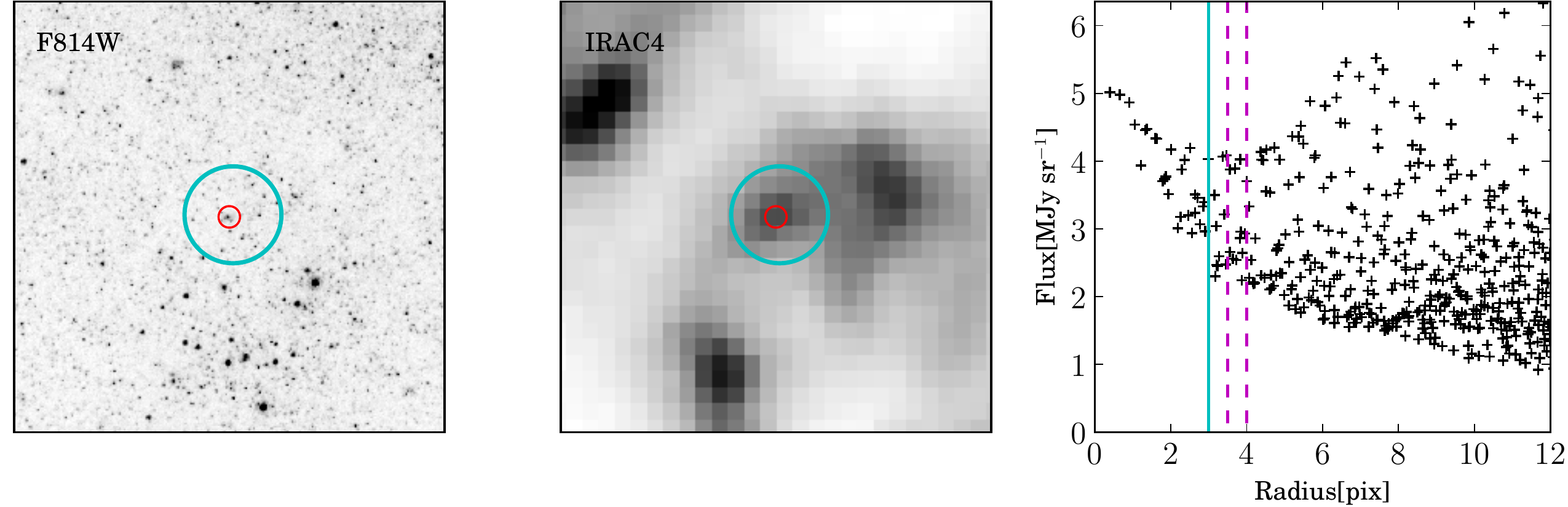}
        \includegraphics[width=\textwidth]{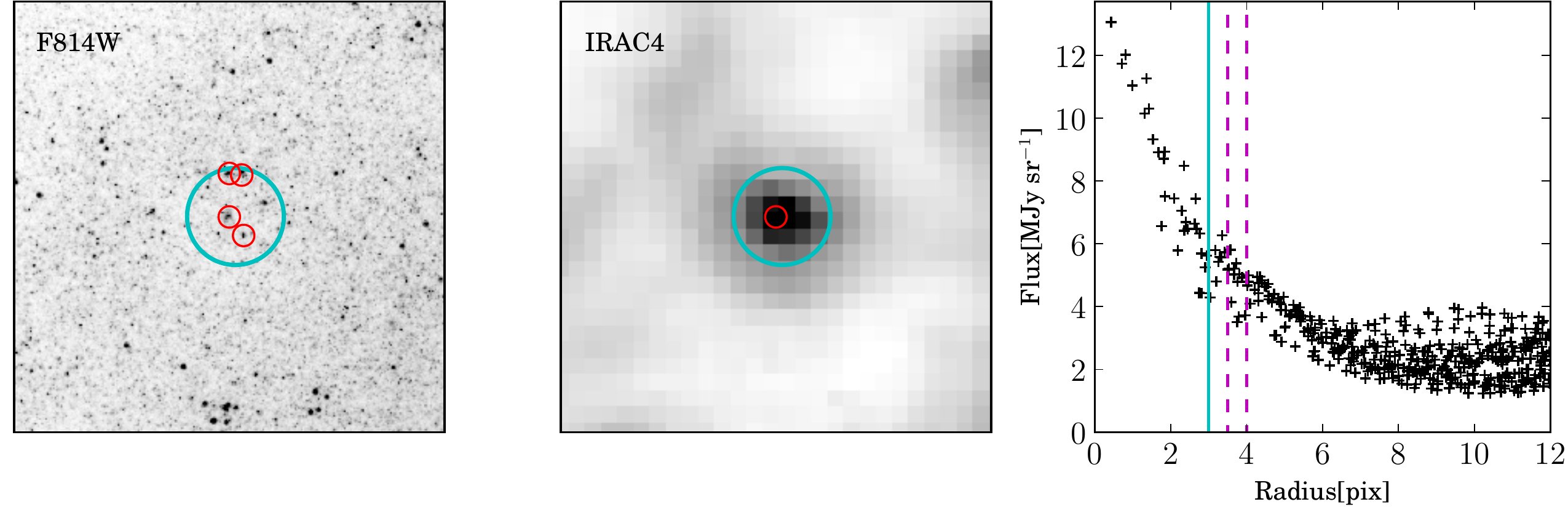}
        \caption{Example of photometry for one cluster (top row) and multiple clusters (bottom row) within one IRAC aperture. For each row, we show the $I$-band (F814W) image (left), IRAC4 image (middle), and the radial profile of the IRAC source (right). In the left and middle panels, the large cyan circles mark the aperture (3 pixel radius) used to perform IRAC4 photometry in this work, while the small red circles locate clusters within one IRAC aperture based on their optical coordinates from the \cite{Adamo2017} catalogs. On the IRAC4 image, only the cluster with the smallest distance to the center of the IRAC4 source is shown. The cyan solid and two magenta dashed lines in the right panel indicate apertures of 3, 3.5, and 4 pixels. Both grayscale images (together with other grayscale images in this paper) are shown in linear scale.
        \label{fig:radprof}}
        \end{figure*}

        We mark the locations of all the clusters on the cutouts of optical and IRAC4 images and inspect each cutout by eye to identify YSCs with IRAC4 counterparts within 2\farcs25.\footnote{This distance is chosen to match the radius of photometric aperture of IRAC4 sources.} For each IRAC4 counterpart, we perform photometry with aperture radii of 3, 3.5, and 4 pixels and sky annuli set at a radius of 10 pixels and a width of 1 pixel. The total fluxes from these three apertures after aperture corrections are in good agreement with each other within the uncertainties. However, we still find that, for some sources, a larger photometric aperture can result in a slightly elevated flux. To avoid any contamination from either neighbors or local background, we adopt 3 pixels (2\farcs25) as the final aperture radius. Additional examination of the radial profiles of each source is carried out by visual inspection, and IRAC4 sources without well-defined profiles within 3 pixels, due to confusion with neighboring sources, are also removed. In Figure \ref{fig:radprof} we show two examples of YSCs with IRAC4 counterparts. The top row has only one YSC within the IRAC4 aperture, while the bottom row is an example of multiple clusters within one aperture. The centers of IRAC4 clumps might have small offsets from the coordinates of YSCs in the \cite{Adamo2017} catalogs even in the case of a single cluster within one IRAC4 aperture. Therefore, we allow the PHOT task to recenter before performing photometry. From the radial profile shown in the top row of Figure \ref{fig:radprof}, it is obvious that adopting a photometric aperture of 4 pixel radius could overestimate the 8 \um\ fluxes due to the contamination of nearby bright sources.

        \begin{figure*}[htb]
        \centering
        \includegraphics[width=\textwidth]{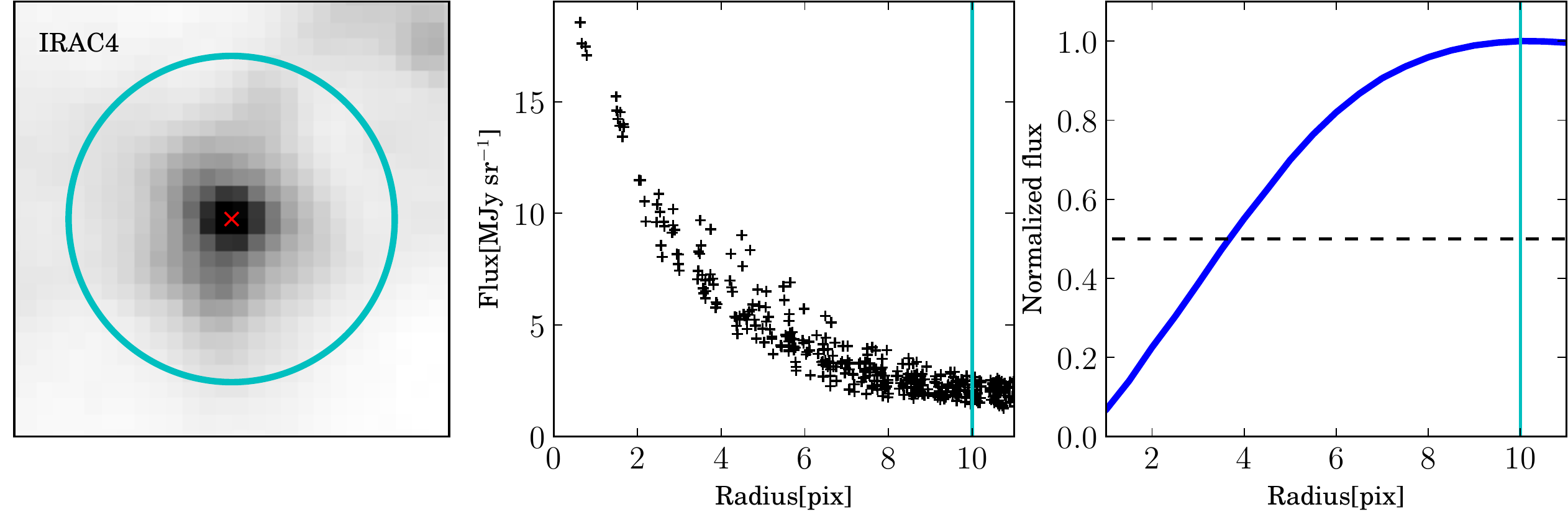}
        \includegraphics[width=0.8\textwidth]{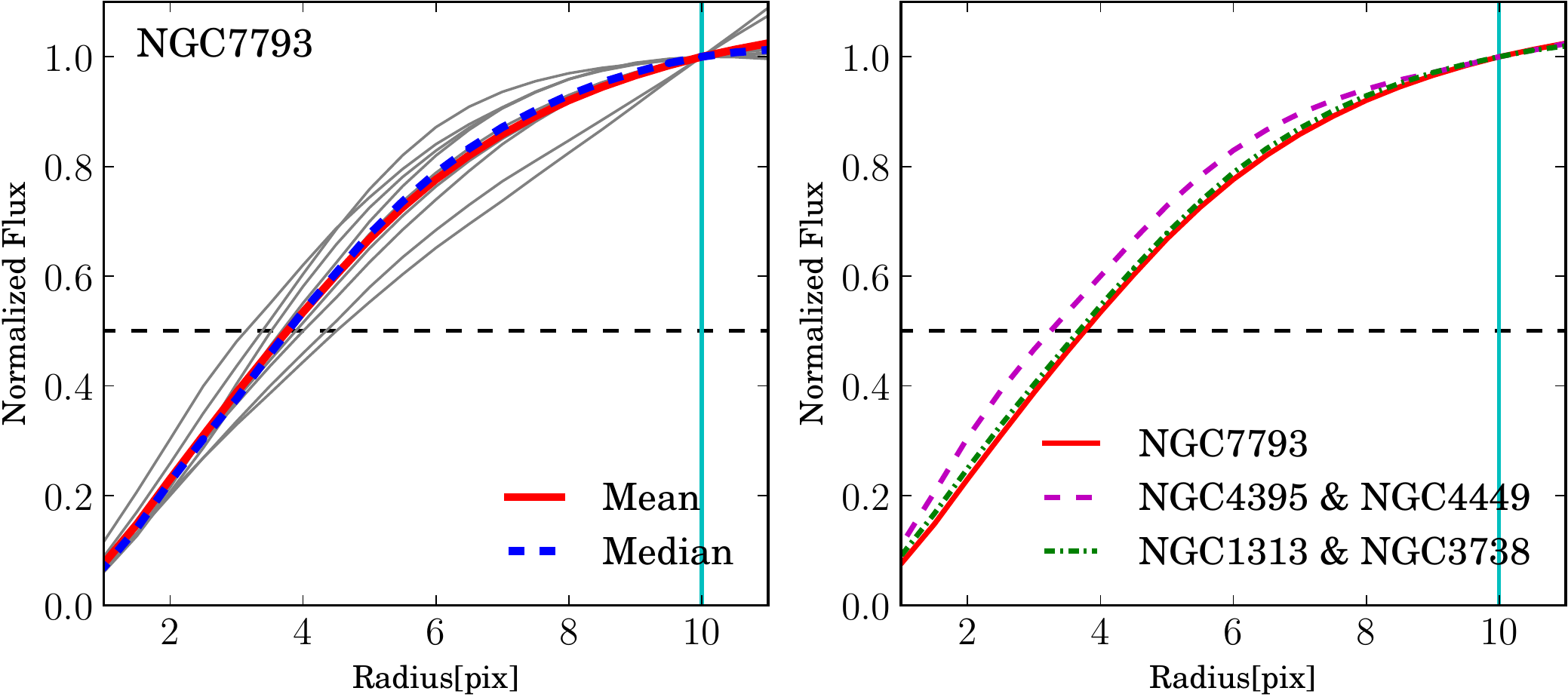}
        \caption{Example of the construction of growth curves for IRAC4 sources. Top row: IRAC4 image (left), radial profile (middle), and growth curve (right) for one reference clump in NGC 7793. The red $\times$ on the IRAC image locates the center of the source. The horizontal black dashed line denotes where the normalized flux = 0.5. Bottom row: the left panel shows all the growth curves (thin gray curves) for NGC 7793 and the mean (red solid one) and median (blue dashed one) curves. The mean curve is adopted as the final growth curve for each galaxy. The right panel presents all the growth curves used in this work. NGC 4395 (NGC 1313) shares the same curve with NGC 4449 (NGC 3738) due to the lack of enough isolated local peaks in both NGC 4449 and NGC 3738. The cyan circle in the top-left panel and cyan vertical lines in the other four panels indicate an aperture of 10 pixels.
        \label{fig:growth_curves}}
        \end{figure*}

        Aperture corrections from our 3 pixel aperture radius to an infinite aperture are derived for our sources as follows. We first identify some isolated local peaks on the IRAC4 images that exhibit clean radial profiles and seem not to be significantly contaminated by nearby sources. Taking these selected sources as reference clumps, we then perform photometry on a set of aperture radii (from 0.5 to 10 pixels with a step of 0.5 pixels) for the selected sources to obtain the growth curves. The sky (local background) is measured within an annulus at 10 pixel radius, 1 pixel wide. After visual inspection of all the radial profiles in the reference sample, this aperture is found to be large enough to enclose most of the flux of isolated sources. Therefore, we adopt fluxes within 10 pixel radius (corresponding to $\sim 145$ pc at a distance of 4 Mpc) as total fluxes for the reference clumps that are used to normalize growth curves. An example of the reference clumps, including the IRAC4 image, the radial profile, and the constructed growth curve, in NGC 7793 is shown in the top row of Figure \ref{fig:growth_curves}. It is evident that this source is isolated enough to obtain a robust growth curve.

        For each galaxy, at least five reference clumps are used to construct the growth curve, and the mean curve for each is adopted as the final curve. In the bottom row of Figure \ref{fig:growth_curves} we present all the growth curves for NGC 7793 (left panel) and all the curves used in this work (right panel). Because the number of isolated local peaks is not sufficiently large to obtain a growth curve for NGC 4449 and NGC 3738, we take the ones from galaxies at similar distances. In other words, NGC 4395 (NGC 1313) shares the same growth curve with NGC 4449 (NGC 3738). The choices of growth curves for NGC 4449 and NGC 3738 only alter the total 8 \um\ fluxes by up to 20\%. Such fluctuations do not significantly change the data and thus the conclusions of this work. For a given aperture, the aperture correction is estimated as the ratio between the fluxes within a 10 pixel radius and the given aperture (3 pixels) for the mean growth curve. The standard deviation of the corrections given by all growth curves within one galaxy is also computed and combined with all other uncertainties.

        The IRAC Instrument Handbook\footnote{\url{https://irsa.ipac.caltech.edu/data/SPITZER/docs/irac/iracinstrumenthandbook/}} specifies that sources smaller than 8\arcsec--9\arcsec\ , like our sources, should be treated as point sources. The IRAC point-source photometry is calibrated on a 12\arcsec\ radius aperture, implying that an additional aperture correction needs to be applied to our 7\farcs5\ aperture photometry. Table~4.7 of the Handbook provides a correction of $\sim$1.05 between a 7\farcs5\ aperture photometry and the default 12\arcsec\ point-source calibration, which we apply to all our 8~$\mu$m measurements.

    \subsection{Mass Correction}
    \label{ssec:mass_correction}

        Given that the 8 \um\ fluxes are measured within an aperture radius of 3 pixels (2\farcs25) on the IRAC4 images, all sources enclosed by this aperture may contribute to the 8 \um\ flux. For this reason, when we consider the total stellar mass corresponding to the measured 8 \um\ fluxes, the mass contribution from the unidentified clusters or individual stars surrounding the identified YSCs should be included.

        We perform photometry on the F814W LEGUS images with the same aperture as the one used for the IRAC4 photometry (i.e., the cyan circles in Figure \ref{fig:radprof}, left) to obtain the total $I$-band flux in the aperture ($f_{I,\mathrm{ap}}$). The $M_*$ and $I$-band fluxes of individual identified YSCs are retrieved from the \cite{Adamo2017} catalogs. We sum the $I$-band fluxes of all identified YSCs within the aperture to get the total flux counted by our selection ($f_{I,\mathrm{sel}}$). Assuming that all the objects (star clusters and individual stars) within one IRAC4 aperture have a similar mass-to-light ratio ($M_*/L_{I}$), the $f_{I,\mathrm{ap}}/f_{I,\mathrm{sel}}$ ratio is applied to the total $M_*$ of the identified YSCs within the aperture as a mass correction. The assumption of a constant mass-to-light ratio within each aperture is equivalent to assuming that the bulk of the $I$-band light originates from same-age stellar populations. This assumption is likely to be acceptable within our small regions ($\lesssim50$ pc). We show in Appendix \ref{appen:new_fit} that our regions are better described by more complex stellar populations than a single-age one; however, the impact on the estimated mass in the IRAC4 aperture is small (see Figure~\ref{fig:comp_m_age}). For all selected IRAC4 clumps, $f_{I,\mathrm{sel}}$ only represents a small fraction of $f_{I,\mathrm{ap}}$, leading to a fairly large mass correction with a median and 68\% range of $4.93_{-2.51}^{+5.89}$.

\section{Results}
\label{sec:results}

    Given our goal of investigating the relation between the ages of YSCs and their PAH emission, we use the mass-weighted age ($\mathrm{age}_{M_*}$) and normalize the 8 \um\ luminosity by the total stellar mass within the IRAC4 aperture ($\nu L_{\nu,8}/M_*$) to avoid biasing our quantities according to the most massive cluster/contribution. $\mathrm{Age}_{M_*}$ is constructed from the values of $M_*$ and the ages of individual clusters in each aperture, as provided in the \cite{Adamo2017} catalogs. In the case where there is only one YSC within the aperture, the measured age of the YSC is directly taken as $\mathrm{age}_{M_*}$. The mass correction is added to the total $M_*$ used in $\nu L_{\nu,8}/M_*$. Due to the asymmetric uncertainties of $M_*$ and age in the catalogs, typical uncertainties are needed to perform error propagation in calculating $\mathrm{age}_{M_*}$ and $\nu L_{\nu,8}/M_*$. For $M_*$, the median and mean of its uncertainty are 18\% and 58\%, respectively. The typical uncertainty of $M_*$ is conservatively assumed to be 60\%. The uncertainties in age have a median and mean of 25\% and 248\%, respectively. Because the mean is biased by several extreme cases, we adopt 25\% as the typical uncertainty in the age.

    \begin{figure*}[htb]
    \centering
    \includegraphics[width=0.48\textwidth]{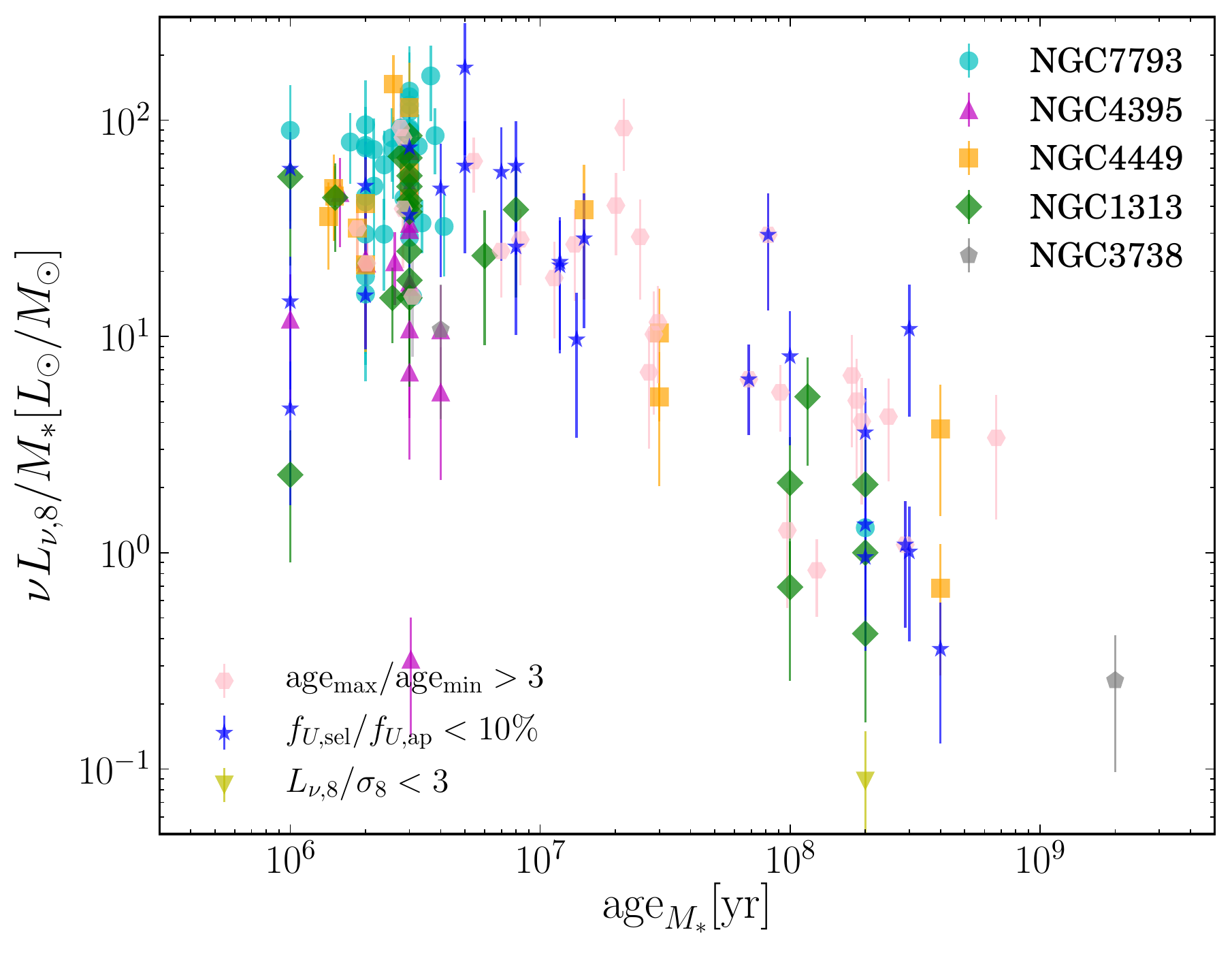}
    \includegraphics[width=0.48\textwidth]{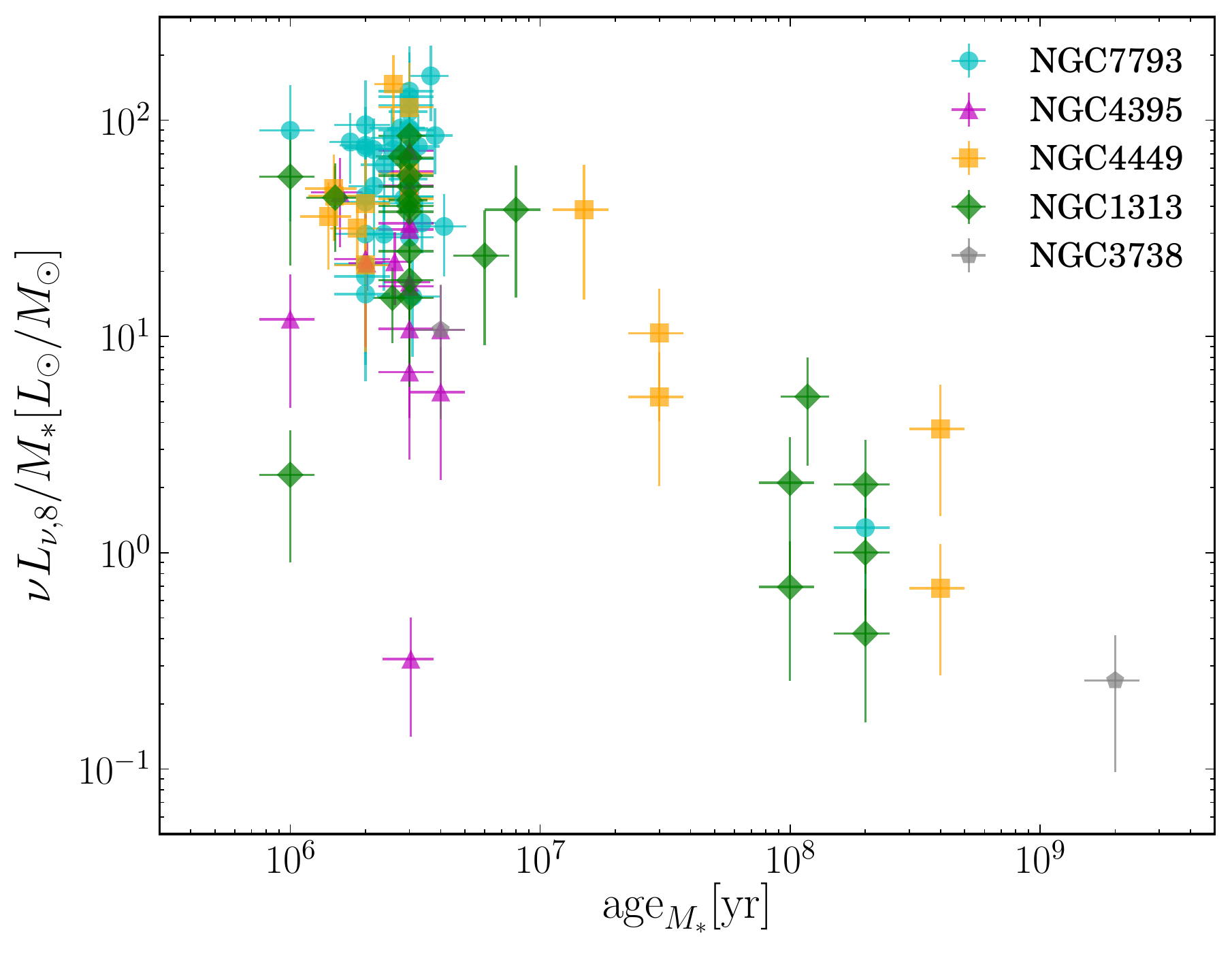}
    \caption{$\nu L_{\nu,8}/M_*$ as a function of $\mathrm{age}_{M_*}$ for all 146 identified IRAC4 counterparts (left) and our final sample (right). The cyan circles, magenta upward-pointing triangles, orange squares, green diamonds, and gray pentagons indicate IRAC4 sources included in our final sample from NGC 7793, NGC 4395, NGC 4449, NGC 1313, and NGC 3738, respectively. In the left panel, the pink hexagons, blue stars, and yellow downward-pointing triangles are sources excluded by three additional criteria; see the text for more details. The uncertainties in $\mathrm{age}_{M_*}$ are not plotted in the left panel in order to show the distribution clearly.
    \label{fig:l8_agem}}
    \end{figure*}

    From the 5 nearby galaxies in our sample, we identify 146 sufficiently isolated IRAC4 regions to perform photometry. Additional criteria are employed to reduce the uncertainties in our final results. First, the mass correction described in Section \ref{ssec:mass_correction} assumes a similar $M_*/L_{I}$ for objects within the same aperture. In the case of YSCs, the assumption means that these clusters should have similar ages. Thus, when multiple clusters are located within one IRAC4 aperture, we require that the largest difference in their ages is not larger than a factor of 3, i.e., $\mathrm{age_{max}/age_{min}}\leq 3$. Second, in some cases, clusters in our catalogs only account for a small fraction of the total flux within the IRAC4 aperture, increasing the risk of adopting an unrepresentative age for the aperture. To decrease the mismatch between the age of the clusters and the age of the population within the IRAC4 aperture, we require the sum of the $U$-band (F336W) fluxes of the YSCs not to be smaller than 10\% of the total flux enclosed by the aperture, i.e., $f_{U,\mathrm{sel}}/f_{U,\mathrm{ap}}\geq 10\%$. Finally, only IRAC4 sources with reliable 8 \um\ detections, i.e., $L_{\nu,8}/\sigma_8\geq 3$, are considered. Here, $\sigma_8$ is the uncertainty of $L_{\nu,8}$ due to local background subtraction and aperture correction. The number of IRAC4 regions that cannot satisfy the above criteria are 24, 27, and 1 for $\mathrm{age_{max}/age_{min}}\leq 3$, $f_{U,\mathrm{sel}}/f_{U,\mathrm{ap}}\geq 10\%$, and $L_{\nu,8}/\sigma_8\geq 3$, respectively. Some of them do not meet more than one criterion. Our final sample includes 97 IRAC4 sources from all 5 galaxies.

    In Figure \ref{fig:l8_agem}, we present the relation between $\nu L_{\nu,8}/M_*$ and $\mathrm{age}_{M_*}$ for all 146 identified IRAC4 sources (left panel) and our final sample (right panel). The numbers of IRAC4 counterparts identified and included in our final sample for each galaxy are listed in Table \ref{tab:properties}. In the left panel, a negative correlation in the $\nu L_{\nu,8}/M_*$ \vs\ $\mathrm{age}_{M_*}$ plane is observed although the scatter around the correlation is fairly large. In the right panel, it is evident that our final sample shows a tight negative relation between $\nu L_{\nu,8}/M_*$ and $\mathrm{age}_{M_*}$, with the exception of two outliers located at the lower-left corner of the panel. To test the robustness of this anticorrelation, we perform a resampling 5000 times with a bootstrap method to calculate the Pearson linear correlation coefficient ($r$). The resulting $r$ has a mean and a standard deviation of -0.77 and 0.08, respectively, indicating a strong and reliable negative linear relation between $\mathrm{age}_{M_*}$ and $\nu L_{\nu,8}/M_*$. The exclusion of the two outliers gives a slightly stronger correlation with $r=-0.84\pm0.05$. Thus, older populations heat the dust (and the PAHs) but result in lower luminosities than younger stellar populations containing massive, UV-bright stars, as may be expected from heuristic considerations. The trend is followed even by the less reliable sources removed from our sample (Figure \ref{fig:l8_agem}, left).

    \begin{figure*}[htb]
    \centering
    \includegraphics[width=0.8\textwidth]{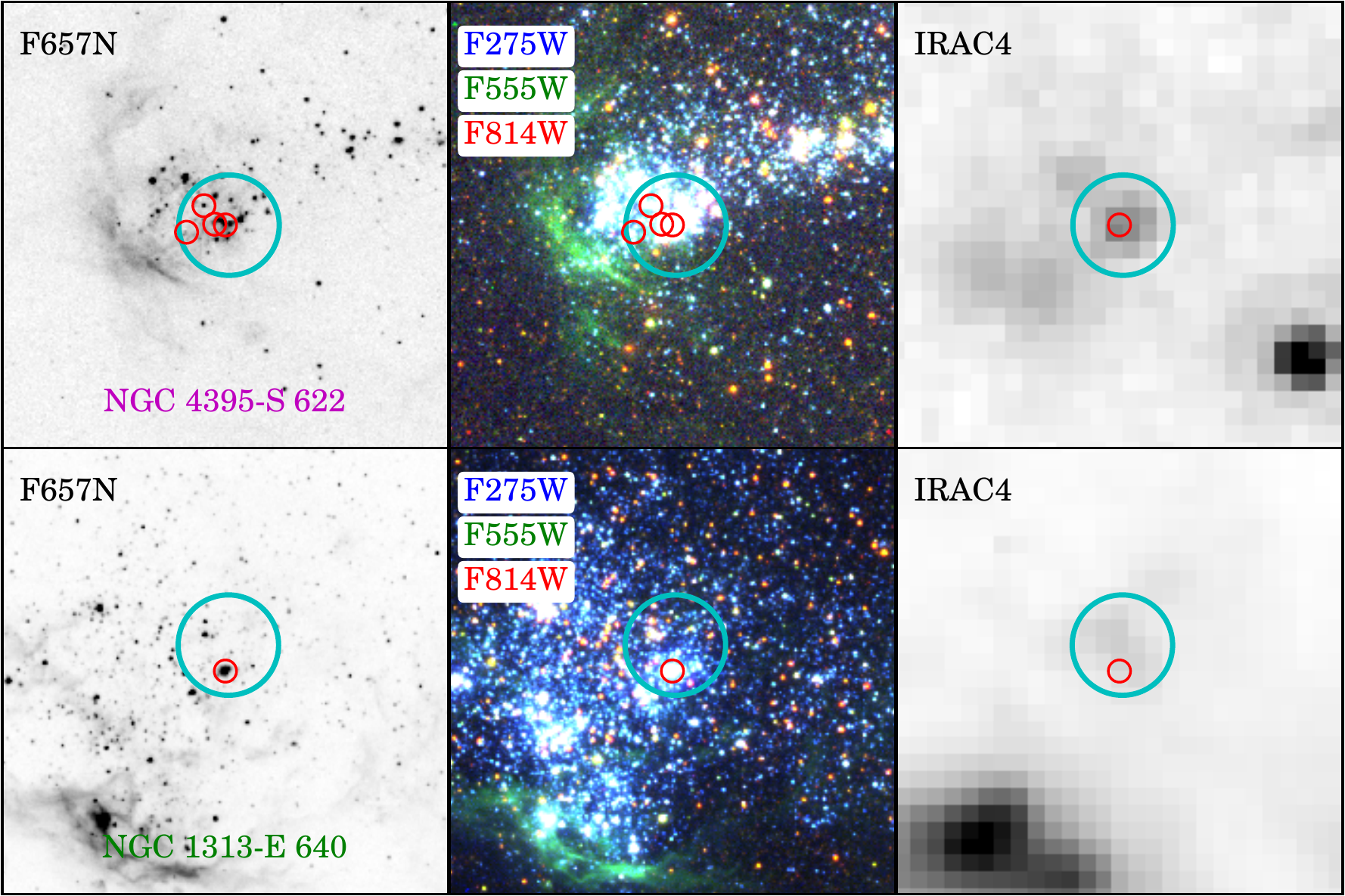}
    \caption{F657N (\ha; left), optical RGB (middle), and IRAC4 (right) images of two outliers in the right panel of Figure \ref{fig:l8_agem}: NGC 4395-S 622 (top) and NGC 1313-E 640 (bottom). NGC 4395-S and NGC 1313-E indicate the pointings, while the following numbers are source IDs of the central clusters in the corresponding \cite{Adamo2017} catalogs. The colors of the source IDs are the same as those of the corresponding points in the right panel of Figure \ref{fig:l8_agem}. The RGB images are composed of the F275W (blue), F555W (green), and F814W (red) images. The small red circles and large cyan circles are the same as Figure \ref{fig:radprof}.
    \label{fig:outliers}}
    \end{figure*}

    The F657N (\ha), optical RGB, and IRAC4 images of the two outliers of Figure \ref{fig:l8_agem}, right, are shown in Figure \ref{fig:outliers}. The F657N (\ha) images of NGC 4395 and NGC 1313 (Program: GO--13773; PI: Rupali Chandar) are retrieved from the Hubble Legacy Archive\footnote{Based on observations made with the NASA/ESA Hubble Space Telescope, and obtained from the HLA, which is a collaboration between the Space Telescope Science Institute (STScI/NASA), the Space Telescope European Coordinating Facility (ST-ECF/ESA), and the Canadian Astronomy Data Centre (CADC/NRC/CSA).} (HLA). The optical RGB images are created by combining F275W (blue), F555W (green), and F814W (red) images. Both outliers are located in crowded regions where the light from the YSCs provides a minor contribution and the other sources within the apertures show a range of colors that may be indicative of a large range of ages. Therefore, the stellar properties of the identified YSCs might be not representative of other stellar objects within the same IRAC4 aperture. Shell-like structures in the \ha\ images and the green channel (F555W) of the optical RGB images also suggest that they are both in regions where the (ionized) gas and dust seem to have been cleared away, resulting in low PAH/dust luminosities. In the following analysis, we will focus on the main sequence formed by the final sample except the outliers.

    Based on observations of $z\sim2$ SFGs, \cite{Shivaei2017} reported an excess of $\nu L_{\nu,8}/L_{\mathrm{TIR}}$ ($L_{7.7}/L_{\mathrm{IR}}$ in their paper) by a factor of $\sim 3$ for galaxies older than 900 Myr compared to younger systems, which was attributed to the delayed injection of carbon dust by AGB stars \citep{Galliano2008}. Due to the lack of $L_{\mathrm{TIR}}$ measurement, we are unable to examine how the $\nu L_{\nu,8}/L_{\mathrm{TIR}}$ varies with age in our sample. However, we note that the reported transition age ($\sim 900$ Myr) is larger than most of the $\mathrm{age}_{M_*}$ of our final sample, suggesting that the local PAH abundance of our sources might not be enriched by the current generation yet, even if the delayed injection of carbon dust indeed exists. Therefore, in the following analysis, we assume an age-independent PAH abundance for our sample.

\section{Comparison with Models}
\label{sec:models}

    To understand the main drivers of the trend and the scatter of the observed $\mathrm{age}_{M_*}$--$\nu L_{\nu,8}/M_*$ correlation, we combine models of stellar populations with models of dust emission through a simple set of assumptions. The Starburst99 stellar population evolution model \citep{Leitherer1999,Leitherer2010,Leitherer2014,Vazquez2005} is used to generate SEDs of YSCs with different ages. We consider an energy-balance argument in which the total luminosity of stellar light absorbed by dust is equal to that of the dust emission in the mid-IR and far-IR. Thus, by assuming a fixed dust absorption fraction ($f_{\mathrm{abs}}$), we integrate the SEDs from UV to NIR (912 \AA--2 \um) to obtain the total IR luminosity, i.e.,
    \begin{equation}\label{eq:LIR}
        L_{\mathrm{TIR}} = f_{\mathrm{abs}}\int_{912\mathrm{\AA}}^{2\mu\mathrm{m}}L_{\lambda}\ \mathrm{d}\lambda,
    \end{equation}
    in which $L_{\lambda}$ represents the SED of the YSC. The \cite{Draine2007} dust models are adopted to calculate the 8 \um\ luminosity fraction from the total IR luminosity ($\nu L_{\nu,8}/L_{\mathrm{TIR}}$). We choose the dust models with size distributions that can reproduce the observed shape of the Milky Way extinction curve but with different PAH abundances ($q_{\mathrm{PAH}}$), i.e., MW3.1\_00 ($q_{\mathrm{PAH}}=0.47\%$), MW3.1\_10 (1.12\%), MW3.1\_20 (1.77\%), MW3.1\_30 (2.50\%), MW3.1\_40 (3.19\%), MW3.1\_50 (3.90\%), and MW3.1\_60 (4.58\%) in \cite{Draine2007}\footnote{The suffixes of model labels represent the total C abundance (per H nucleus) in ppm, e.g., MW3.1\_60 means 60 ppm C/H in PAHs in this model \citep{Weingartner2001}. Since the publication of the \cite{Draine2007} model, observations have suggested that the dust/gas ratio used in this model is systematically overestimated \citep{Aniano2020}. \cite{Aniano2020} found that a typical correction of $\sim0.62$ should be applied to dust mass derived from the \cite{Draine2007} model for their sample. Thus, the conversion from carbon abundance (in ppm) to $q_{\mathrm{PAH}}$ should be updated. This will change the C abundance for a given \cite{Draine2007} model, while the $q_{\mathrm{PAH}}$ is unaffected because it is a relative value in mass (B. T. Draine 2020, private communication). For instance, the model with $q_{\mathrm{PAH}}=4.58\%$ should correspond to a C abundance of $\sim0.62\times60~\mathrm{ppm} = 37$ ppm C/H in PAHs. Therefore, we caution that model labels like ``MW3.1\_60'' quoted here should be only treated as labels.}. The $q_{\mathrm{PAH}}$ is defined as the mass fraction of the dust grains in the form of PAHs containing fewer than 1000 carbon atoms. These models are parameterized by a dimensionless scaling factor $U$, which describes the energy density of the local radiation field normalized by the interstellar radiation field of the solar neighborhood given by \cite{Mathis1983}.

    \begin{figure}[htb]
    \centering
    \includegraphics[width=0.49\textwidth]{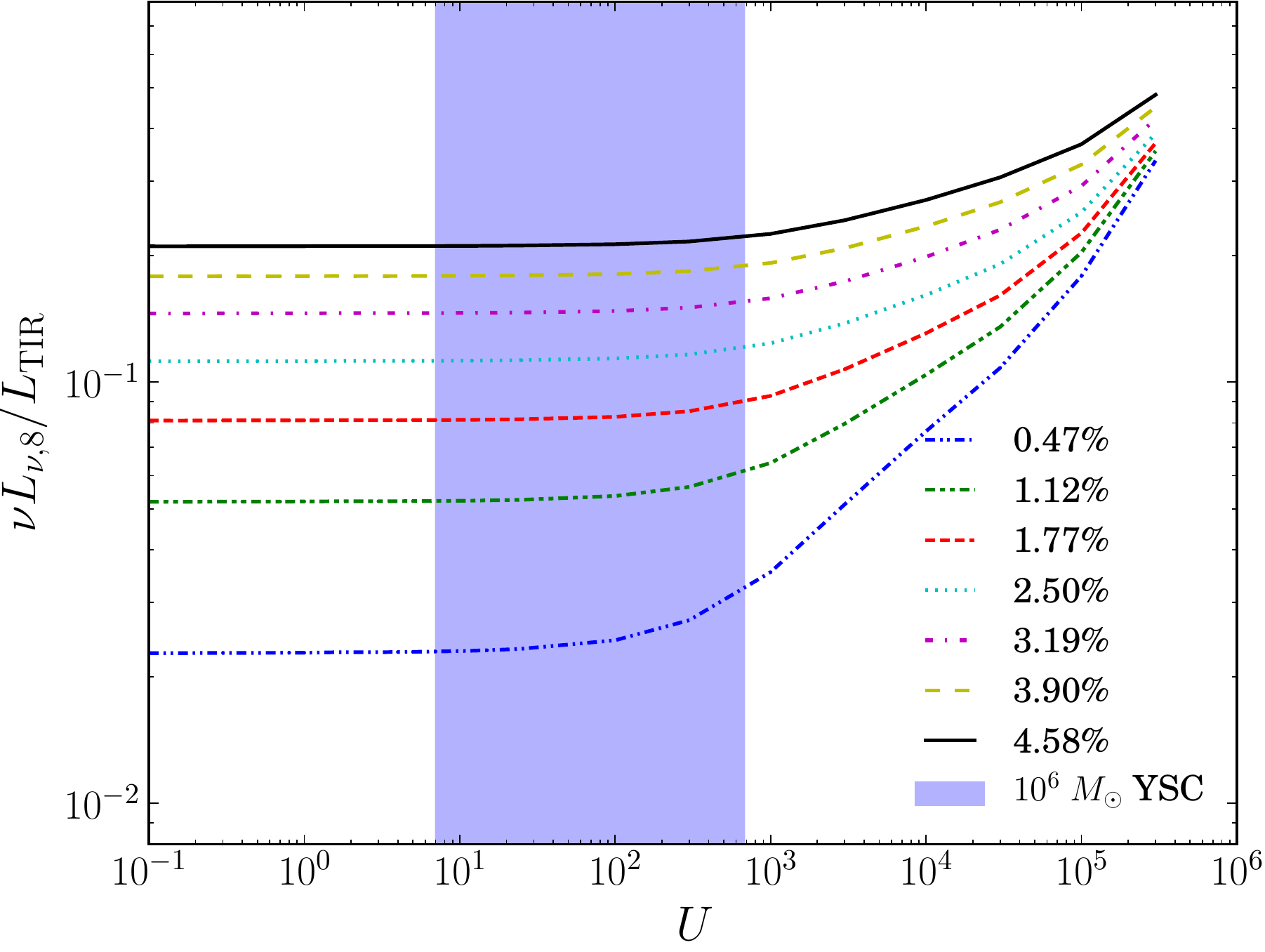}
    \caption{$\nu L_{\nu,8}/L_{\mathrm{TIR}}$ as a function of $U$ for dust models with different $q_{\mathrm{PAH}}$, adapted from \cite{Draine2007}. From bottom to top, the curves represent models with increasing $q_{\mathrm{PAH}}$. The blue shadow region denotes the $U$ range of a PDR excited by a $10^6~M_{\odot}$ star cluster, sampling an age range of 1--100 Myr.
    \label{fig:L8toLIR}}
    \end{figure}

    We plot the $\nu L_{\nu,8}/L_{\mathrm{TIR}}$ ratios as a function of $U$ for dust models with different $q_\mathrm{PAH}$ in Figure \ref{fig:L8toLIR}. Assuming a PDR distance of $\sim45$ pc to the central star cluster, which is estimated from the distance between the peaks of FUV and 8 \um\ emission of a luminous \hii\ region with $M_*\sim10^5~M_{\odot}$ (NGC 595) in M33 \citep{Relano2009}, we create SEDs from Starburst99 for a stellar object with $M_*=10^6~M_{\odot}$ and an age in the range 1--100 Myr, and calculate the $U$ range of its surrounding PDR. As shown in Figure \ref{fig:L8toLIR}, the $U$ range of this massive object covers the flat region of the $U$--$\nu L_{\nu,8}/L_{\mathrm{TIR}}$ relation at which the PAH is entirely excited by single-photon heating and thus the $\nu L_{\nu,8}/L_{\mathrm{TIR}}$ is independent of $U$ \citep{Draine2007}. Therefore, we assume a constant value of $U=12$ in our models. This assumption is consistent with the estimated range of the mean $U$ for local galaxies derived via SED fitting with observations from NIR to far-IR (FIR) and the \cite{Draine2007} dust model \citep{Draine2007a}. As a result, the $\nu L_{\nu,8}/L_{\mathrm{TIR}}$ is fixed for each dust model and independent of the age of the YSCs.

    \subsection{Variations in SSP Models}
    \label{ssec:vary_ssp}

        We first examine how the assumed simple stellar population (SSP) models affect the PAH emission. The ingredients in the SSP models considered here are the stellar evolutionary tracks and metallicities, while the IMF is fixed to the \cite{Kroupa2001} one. We adopt five tracks for the Starburst99 models we generate: GENEVA STD, GENEVA HIGH, PADOVA STD, PADOVA AGB, and GENEVA V00. The first and second tracks are the Geneva models with standard \citep{Schaller1992,Schaerer1993,Schaerer1993a,Charbonnel1993} and high \citep{Meynet1994} mass-loss rates, respectively. The third and fourth tracks are the original Padova models \citep{Bressan1993,Fagotto1994,Fagotto1994a,Girardi2000} and the ones with thermally pulsing AGB stars \citep{Vazquez2005}, respectively. The last one is the updated Geneva models with zero rotation \citep{Ekstroem2012,Georgy2013,Leitherer2014}. More details can be found in the online document\footnote{\url{http://www.stsci.edu/science/starburst99/docs/run.html}} or the relevant literatures \citep{Vazquez2005,Leitherer2014}. Given that the metallicities of four out of five selected galaxies are about half of the solar metallicity (see Table \ref{tab:properties}), we only consider three metallicities around this value: $Z=0.004$, 0.008, 0.02.

        \begin{figure}[htb]
        \centering
        \includegraphics[width=0.48\textwidth]{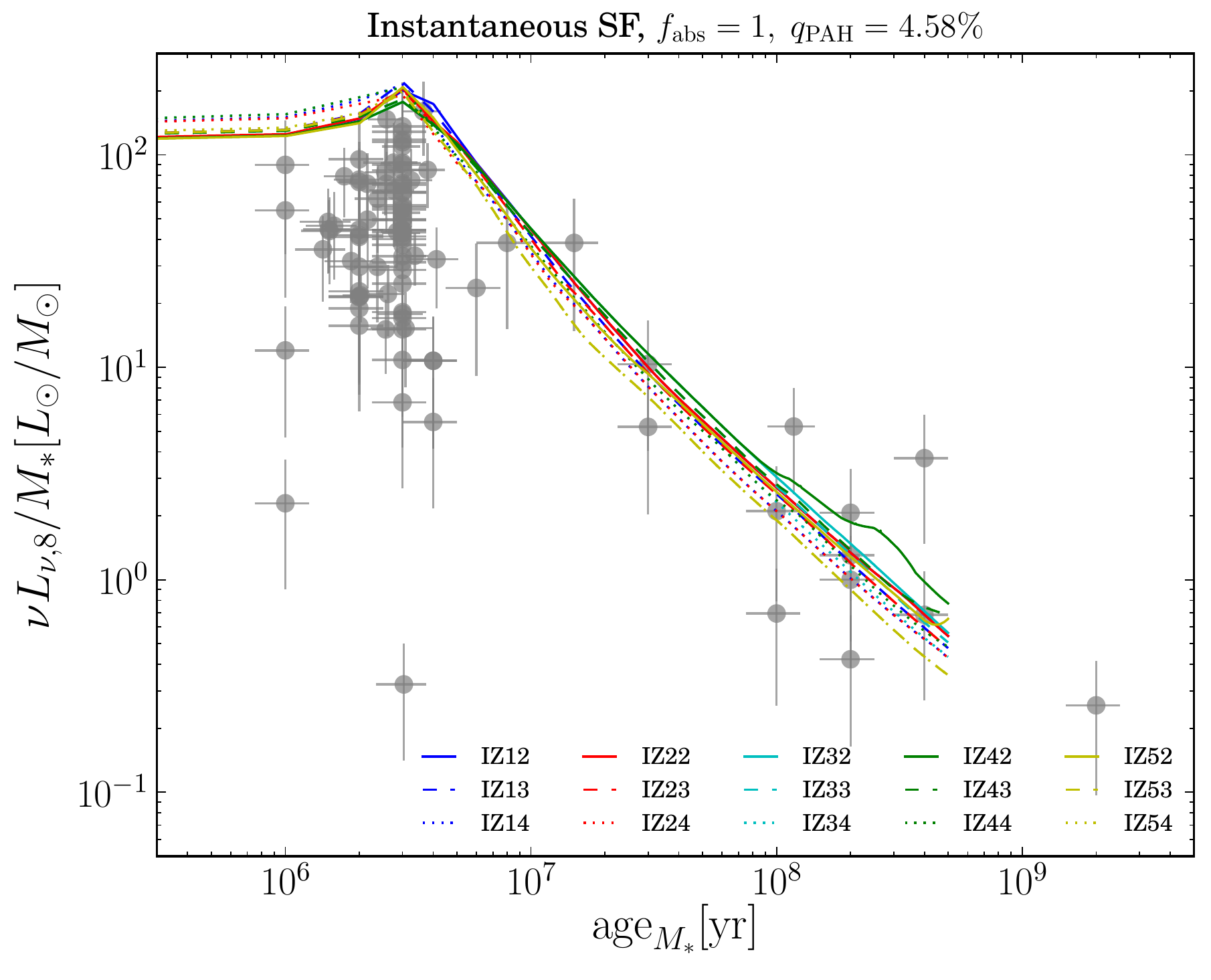}
        \caption{$\nu L_{\nu,8}/M_*$ as a function of $\mathrm{age}_{M_*}$ for instantaneous star formation constructed from different SSP models. The underlying gray circles with error bars are the same as the data shown in the right panel of Figure \ref{fig:l8_agem}. The adopted SSP models are taken from Starburst99 and labeled as IZxy, where $x$ indicates the stellar evolutionary tracks: 1 -- GENEVA STD, 2 -- GENEVA HIGH, 3 -- PADOVA STD, 4 -- PADOVA AGB, 5 -- GENEVA V00; and $y$ indicates the metallicity: 2 -- 0.004 (except IZ52 for which $Z = 0.002$), 3 -- 0.008, 4 -- 0.02 (except IZ54 for which $Z = 0.014$).
        \label{fig:vary_ssp}}
        \end{figure}

        Adopting a dust model with MW3.1\_60 ($q_{\mathrm{PAH}}=4.58\%$), $f_{\mathrm{abs}}=1$, and instantaneous star formation, we run Starburst99 from 0.01 to 500 Myr. $f_{\mathrm{abs}}=1$ means that all the optical light is absorbed by dust. This assumption would not be sensible given that clusters in our sample are observed optically. However, we here aim to maximize the PAH emission of each SSP model and thus adopt $f_{\mathrm{abs}}=1$ and the highest PAH fraction in the \cite{Draine2007} dust model. We will show later (Section \ref{ssec:vary_dust} and Section \ref{ssec:vary_fabs}) that varying the $q_{\mathrm{PAH}}$ and $f_{\mathrm{abs}}$ values chosen will only change the normalization but not the shape of the relation derived from the models. The predicted relations between $\nu L_{\nu,8}/M_*$ and the ages for different SSP models are shown in Figure \ref{fig:vary_ssp}. For a fixed SSP model, the PAH emission increases slowly at an $\mathrm{age}_{M_*}$ of $\sim 1$ Myr, reaches its maximum at an $\mathrm{age}_{M_*}$ of $\sim 3$ Myr, and then decreases rapidly toward the high-$\mathrm{age}_{M_*}$ end. It can be seen from Figure \ref{fig:vary_ssp} that changing either the evolutionary track or the metallicity of the SSP models does not alter the normalized PAH emission significantly. Note that there are numerous young sources laying below the (maximal) model predictions plotted in Figure \ref{fig:vary_ssp}, we will discuss this scatter in Section \ref{ssec:vary_dust} and Section \ref{ssec:vary_fabs}.

        Previous works have demonstrated that there is a positive correlation between the PAH emission/abundance and metallicity (e.g., \citealt{Engelbracht2005,Madden2006,Draine2007a,Smith2007,Remy-Ruyer2015}). Although we change the metallicity of our tracks, we do not include the resulting metallicity-dependent PAH abundance, which will be examined in Section \ref{ssec:vary_dust}. The conclusion we can draw from Figure \ref{fig:vary_ssp} is that the change in metallicity in the SSP model does not significantly alter the PAH emission in term of input energy.

    \subsection{Variations in SFH}
    \label{ssec:vary_sfh}

        Due to the fact that the aperture used for IRAC4 photometry (3 pixels or 2\farcs25) always encloses more than one stellar object, the observed PAH emission might arise from YSCs with different ages. Although we have applied a criterion that limits the age difference of the YSCs ($\mathrm{age_{max}/age_{min}}\leq 3$), it is still possible that $L_{\nu,8}$ receives additional contributions from young stellar objects that are not in our sample due to their UV-to-$I$ photometry (i.e., detections in less than four bands) or unacceptable SED fitting results ($Q\leq0.001$). Thus, it is reasonable to use complex stellar populations to model the observations, especially for the old objects that lie above all the model predictions plotted in Figure \ref{fig:vary_ssp}.

        \begin{figure}[htb]
        \centering
        \includegraphics[width=0.48\textwidth]{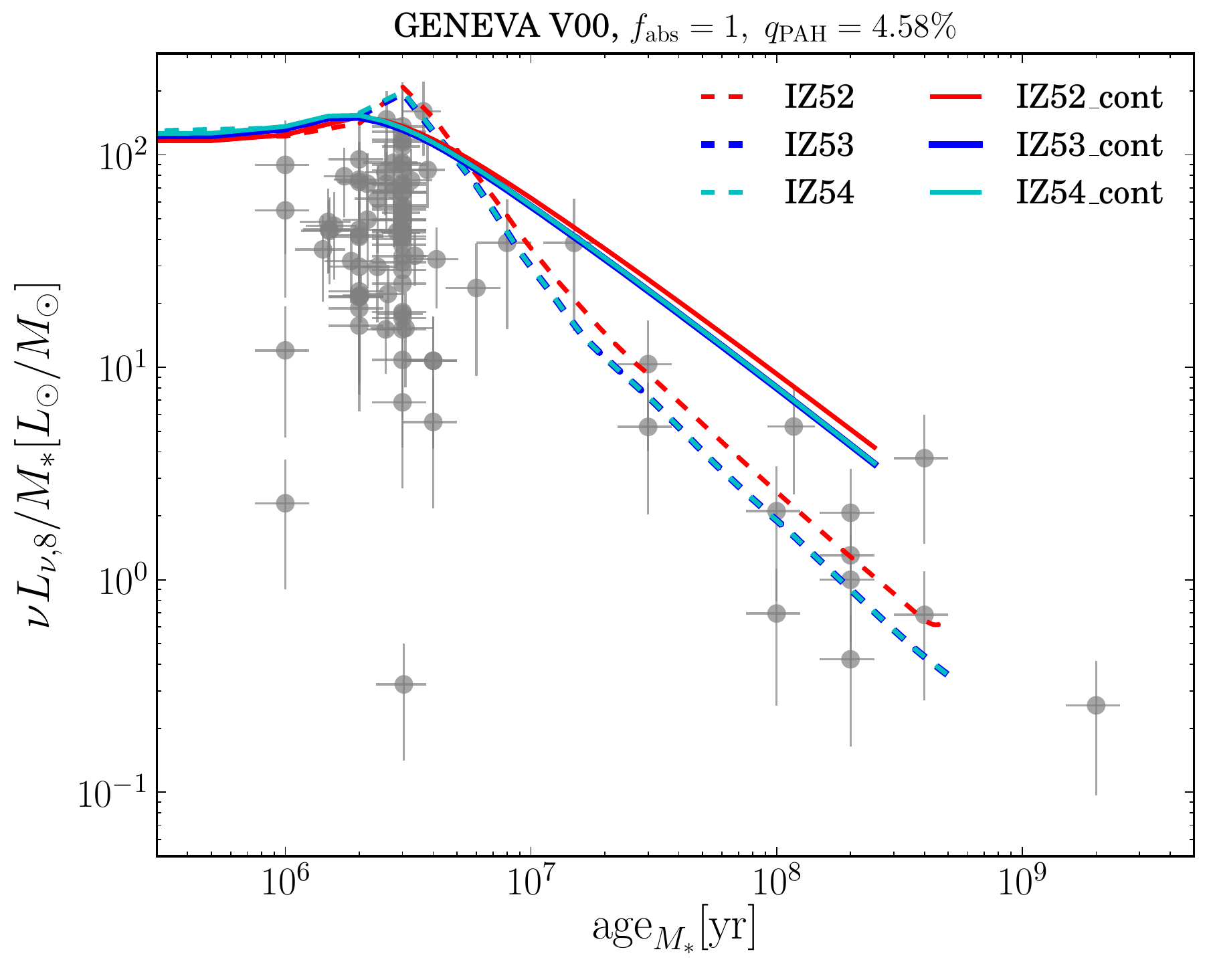}
        \caption{$\nu L_{\nu,8}/M_*$ as a function of $\mathrm{age}_{M_*}$ for different SFHs. The underlying gray circles with error bars are the same as the data shown in the right panel of Figure \ref{fig:l8_agem}. The adopted SSP models are a combination of the Geneva tracks with zero rotation and $Z=0.002$, 0.008, and 0.014, i.e., IZ52, IZ53, and IZ54, respectively. The dashed curves indicate model predictions from instantaneous star formation, while the solid curves are derived based on continuous star formation. Due to the similarity between the IZ53 and IZ54 models, the blue curves are hidden underneath the cyan curves.
        \label{fig:vary_sfr}}
        \end{figure}

        For simplicity, we consider a continuous star formation model with an SFR of 1 $M_{\odot}$ yr$^{-1}$. It is important to note that the predictions of the $\mathrm{age}_{M_*}$--$\nu L_{\nu,8}/M_*$ relation are independent of the SFR assumed here, because $\mathrm{age}_{M_*}$ is weighted by $M_*$ and the PAH emission is normalized by $M_*$. The Geneva tracks with zero rotation (GENEVA V00 in Starburst99) are adopted as our default tracks. Three metallicities ($Z = 0.002$, 0.008, 0.014) are used to run the code. Again, a dust model of MW3.1\_60 ($q_{\mathrm{PAH}}=4.58\%$) and a dust absorption fraction of $f_{\mathrm{abs}}=1$ are assumed when computing $\nu L_{\nu,8}$ in order to minimize the number of free parameters. In Figure \ref{fig:vary_sfr}, we show the comparison of the predicted PAH emission for different star formation histories (SFHs).

        As concluded in the previous section, the assumed metallicity in the stellar SED does not have a strong effect on the output total UV-to-NIR luminosity, and thus the PAH emission. For young objects with an $\mathrm{age}_{M_*}\lesssim 5$ Myr, the strength of the PAH emission does not change with the SFH. However, for an $\mathrm{age}_{M_*}\gtrsim 5$ Myr, a continuous star formation model can greatly enhance the PAH emission, and this enhancement increases with the increasing age of the population. In case of multiple star clusters within one aperture, we find that, at fixed $\mathrm{age}_{M_*}$, the total PAH emission of star clusters with different ages tends to be stronger than that of clusters with similar ages, especially for large $\mathrm{age}_{M_*}$. This trend is also suggested by the sources with $\mathrm{age_{max}/age_{min}}> 3$ (pink hexagons) in the left panel of Figure \ref{fig:l8_agem}.

    \subsection{Variations in Dust Models}
    \label{ssec:vary_dust}

        In the previous subsections, we only consider the dust model with $q_{\mathrm{PAH}}=4.58\%$ and find that the predicted PAH luminosity only matches a small part of the observations. As shown in Figure \ref{fig:L8toLIR}, varying the $q_{\mathrm{PAH}}$ can significantly change the $\nu L_{\nu,8}/L_{\mathrm{TIR}}$ and thus the normalized 8 \um\ luminosity. Here, we go further to see how the selection of dust models (i.e., $q_{\mathrm{PAH}}$) influences the $\mathrm{age}_{M_*}$--$\nu L_{\nu,8}/M_*$ relation.

        \begin{figure}[htb]
        \centering
        \includegraphics[width=0.48\textwidth]{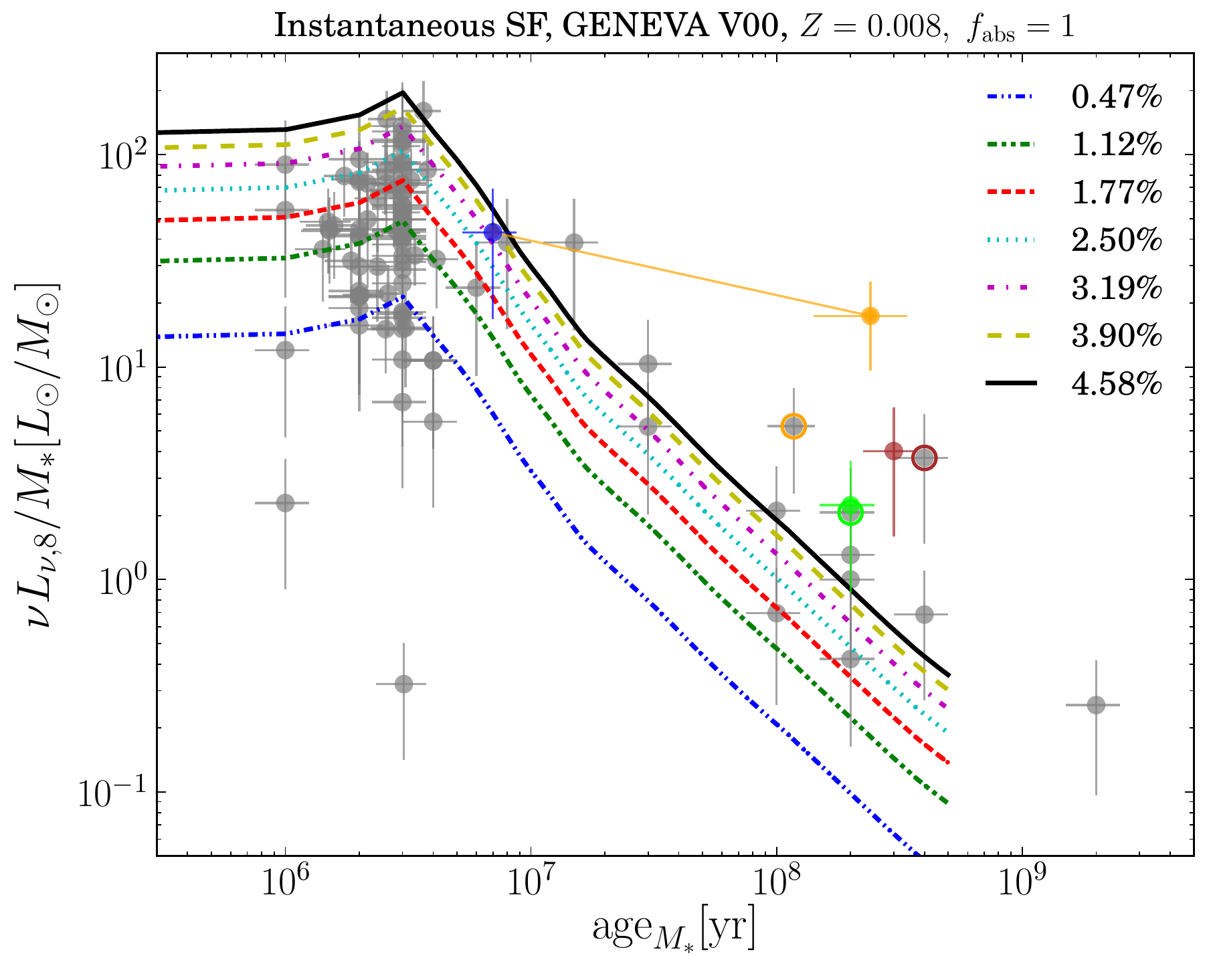}
        \includegraphics[width=0.48\textwidth]{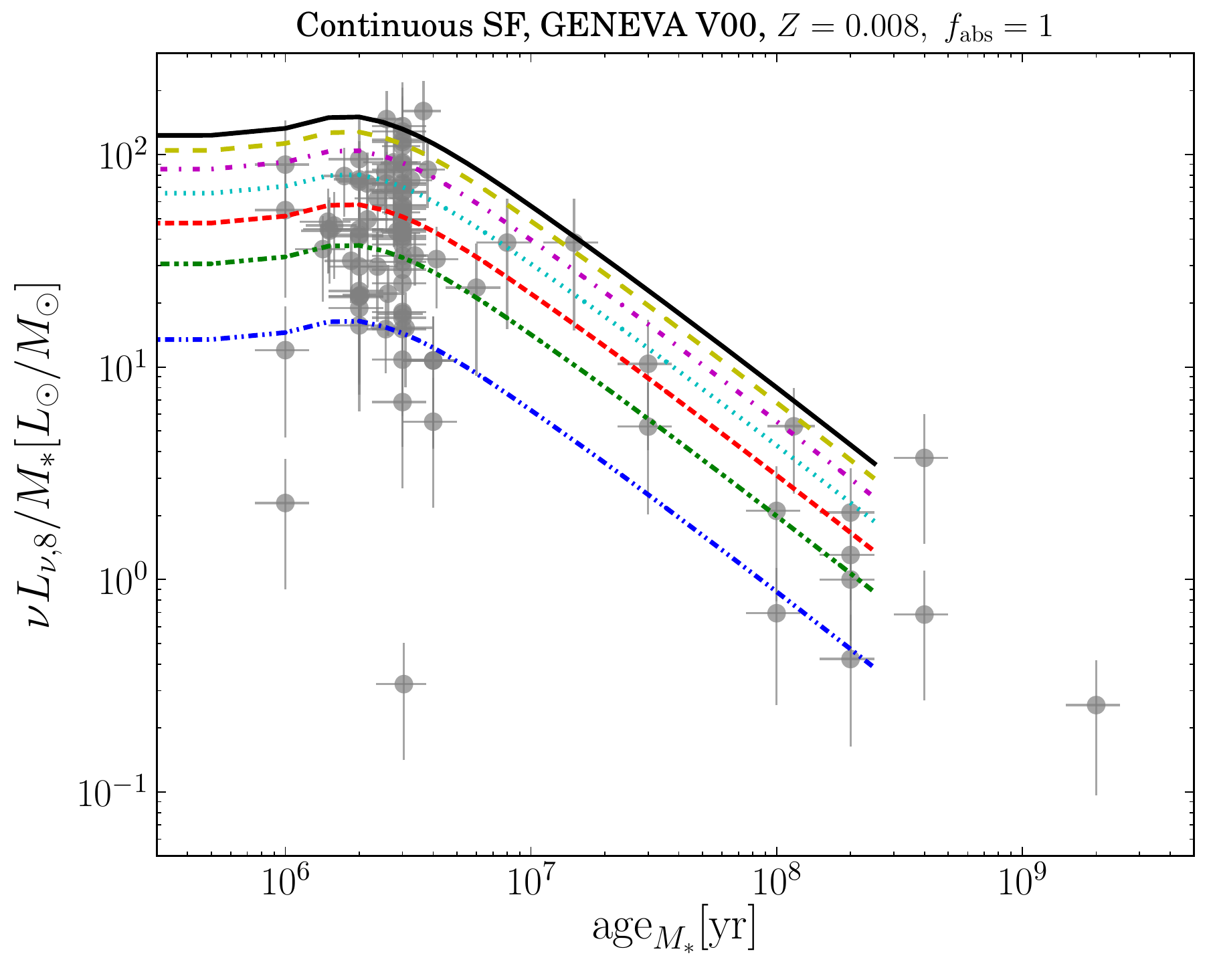}
        \caption{$\nu L_{\nu,8}/M_*$ as a function of $\mathrm{age}_{M_*}$ derived from instantaneous (top) and continuous (bottom) star formation for different dust models. The underlying gray circles with error bars are the same as the data shown in the right panel of Figure \ref{fig:l8_agem}. The adopted SSP models are a combination of the Geneva tracks with zero rotation and $Z=0.008$ (IZ53). In each case, from bottom to top, the curves correspond to dust models with increasing $q_{\mathrm{PAH}}$. In the top panel, the orange, lime, and brown open circles indicate three sources located above the model predictions, while the filled circles with error bars show the locations of the same sources after their SEDs have been fitted with the addition of the narrowband \ha\ fluxes. The blue filled circle with an error bar denotes the location for the aperture of NGC 1313-W 1871 if only the younger cluster (NGC 1313-W 1894, marked as a blue circle in Figure \ref{fig:beyond_singleSF}) is considered. The orange solid line connects the blue filled circle with the orange filled circle, indicating that the latter one is the parent of the former one. See Section \ref{ssec:pah_excess} for further discussion.
        \label{fig:vary_dust}}
        \end{figure}

        Assuming our fiducial Geneva tracks with zero rotation and $Z=0.008$, we run Starburst99 with both instantaneous and continuous ($\mathrm{SFR}=1~M_{\odot}~\mathrm{yr}^{-1}$) star formation. The model predictions for these two SFHs are shown in Figure \ref{fig:vary_dust}. The change in $q_{\mathrm{PAH}}$ from 0.47\% to 4.58\% elevates $\nu L_{\nu,8}/M_*$ by an order of magnitude as expected. What is not expected is that the range of $q_{\mathrm{PAH}}$ brackets the entire scatter observed in the relation between $\nu L_{\nu,8}/M_*$ and $\mathrm{age}_{M_*}$. \cite{Remy-Ruyer2015} indicated that a global $q_{\mathrm{PAH}}$ of $4.58\%$ (the Milky Way PAH abundance) is suitable for these galaxies with subsolar metallicity. Particularly, for NGC 7793, $q_{\mathrm{PAH}}=4.30_{-0.37}^{+0.32}$ was derived in \cite{Remy-Ruyer2015}, while for NGC 4449, the global $q_{\mathrm{PAH}}$ is $\sim2\%$, $3.02_{-0.09}^{+0.14}$, and $\sim3.2\%$ from \cite{Karczewski2013}, \cite{Remy-Ruyer2015}, and \cite{Calzetti2018}, respectively.

        For the instantaneous starburst case, the observations of the youngest populations ($\mathrm{age}_{M_*}\lesssim 10$ Myr) can be explained by models with various PAH abundances, whereas for older populations ($\mathrm{age}_{M_*}\sim 100$ Myr), some sources exhibit a significant enhancement of PAH emission compared to the model predictions. For the continuous star formation, the predicted $\nu L_{\nu,8}/M_*$ is in good agreement with the observations up to an $\mathrm{age}_{M_*}$ of $\sim200$ Myr. These results suggest that additional heating sources of PAH emission enclosed by the IRAC4 aperture might be missed by our YSC selection, especially for sources with large $\mathrm{age}_{M_*}$. Further discussion about sources with excess PAH emission is presented in Section \ref{ssec:pah_excess}.

    \subsection{Variations in Dust Absorption Fraction}
    \label{ssec:vary_fabs}

\begin{figure}[htb]
        \centering
        \includegraphics[width=0.48\textwidth]{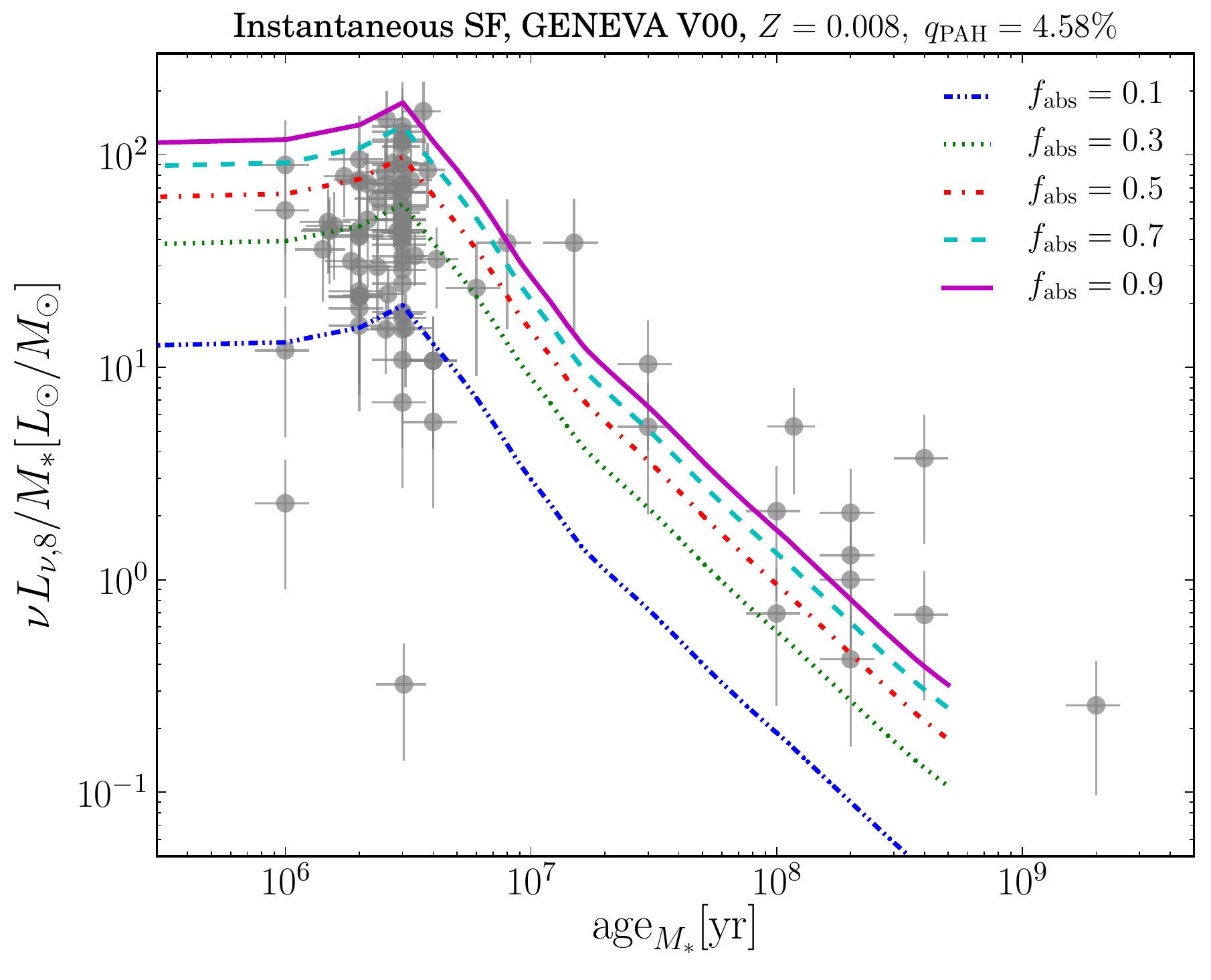}
        \includegraphics[width=0.48\textwidth]{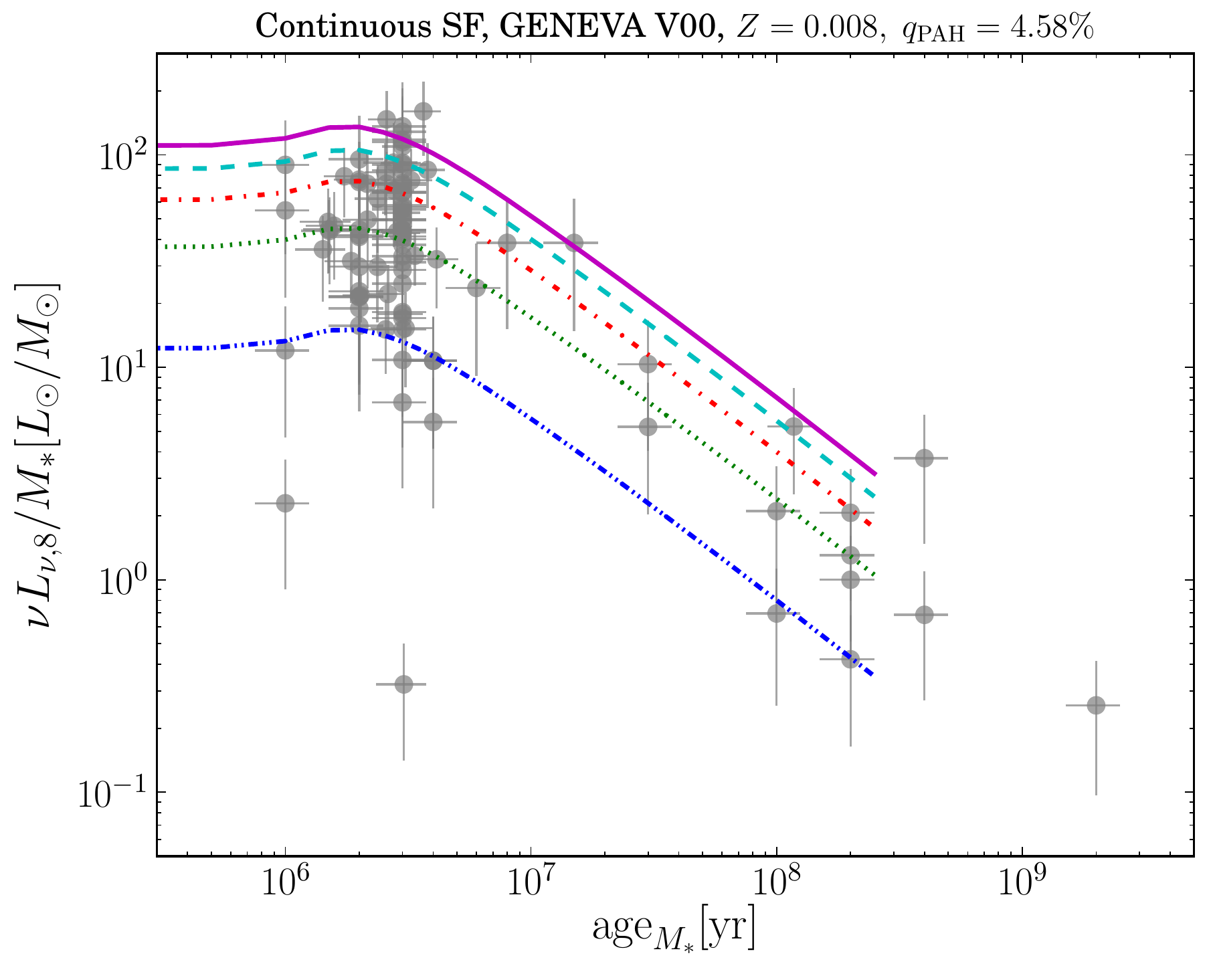}
        \caption{$\nu L_{\nu,8}/M_*$ as a function of $\mathrm{age}_{M_*}$ for different dust absorption fractions $f_{\mathrm{abs}}$. The underlying gray circles with error bars are the same as the data shown in the right panel of Figure \ref{fig:l8_agem}. The adopted SSP models are a combination of the Geneva tracks with zero rotation and $Z=0.008$ (IZ53), together with instantaneous (top) and continuous (bottom) star formation and a dust model with $q_{\mathrm{PAH}}=4.58\%$. In each panel, from bottom to top, the curves indicate model predictions of increasing $f_{\mathrm{abs}}$ from 0.1 to 0.9.
        \label{fig:vary_fabs}}
        \end{figure}

        Finally, we fix the SSP model (GENEVA V00 and $Z=0.008$), the SFH (instantaneous star formation and continuous star formation with SFR = 1 $M_{\odot}$ yr$^{-1}$), and the dust model ($q_{\mathrm{PAH}}=4.58\%$), and vary $f_{\mathrm{abs}}$ from 0.1 to 0.9 (Figure \ref{fig:vary_fabs}). Here, $f_\mathrm{abs}$ is the fraction of UV-to-NIR stellar light absorbed by dust and reemitted in the IR, as indicated in Equation (\ref{eq:LIR}). Similar to the variations in the dust models, the predicted PAH emission reproduces the observed scatter in the data, while the effect of increasing $f_{\mathrm{abs}}$ is very similar to that of increasing $q_{\mathrm{PAH}}$.

        Because we have assumed a fixed $U$ for a given dust model, a constant $\nu L_{\nu,8}/L_{\mathrm{TIR}}$ is assigned to each dust model. Therefore, the role of $f_{\mathrm{abs}}$ in determining $L_{\nu,8}$ is the same as that of $q_{\mathrm{PAH}}$ although they occur under different conditions. In other words, in the current models, $f_{\mathrm{abs}}$ is completely degenerate with $q_{\mathrm{PAH}}$. However, given the small range of metallicity variation in our sample of five galaxies, and recalling that the PAH emission is approximately proportional to the metallicity \citep{Remy-Ruyer2015}, we should expect a small variation in $q_{\mathrm{PAH}}$. Thus, we tend to favor variations in $f_{\mathrm{abs}}$ as the main reason for the observed scatter in $\nu L_{\nu,8}/M_*$ at fixed $\mathrm{age}_{M_*}$.

        Both Figure\ \ref{fig:vary_dust} and Figure \ref{fig:vary_fabs} suggest that our observations are better described by an instantaneous burst model for ages below $\sim$10$^7$~yr, and by continuous star formation for ages above $\sim$10$^7$~yr,
        after including the effects of varying $q_{\mathrm{PAH}}$ and/or $f_{\mathrm{abs}}$. This result is supported by SFH studies based on color-magnitude diagrams. NGC 4449 and NGC 3738 are found to have approximately constant SFHs in the last 200 Myr \citep{Cignoni2018,Cignoni2019,Sacchi2018}. \cite{Sacchi2019} also reported that the SFH of NGC 7793, which contributes 40\% of our final sample (see Table \ref{tab:properties}), is not bursty and suggests inside-out star formation in this galaxy. Further discussion can be found in Appendix \ref{appen:new_fit} where we demonstrate that complex stellar populations are still needed to explain the observations of some of our sources.

    \subsection{Variations in the Hardness of the Local Radiation Field}
    \label{ssec:vary_hardness}

        The \cite{Draine2007} dust model assumes a fixed spectral shape for the local radiation field (i.e., the shape is that of the solar neighborhood, estimated by \citealt{Mathis1983}) and ignores its variations. As the stellar populations age, the extant model only considers the decrease in the amount of input energy, but ignores the variations in the hardness of the local radiation field. The radiative cross section of the PAHs (e.g., Figure 1 in \citealt{Draine2011a}) is wavelength dependent and increases toward the UV, which means that for YSCs with massive OB stars, and thus more UV photons, the absorption/emission of the PAHs will be enhanced. Because the effect on the absorption is included in $f_{\mathrm{abs}}$ and discussed above, here we only consider the effect on the emission in terms of $\nu L_{\nu,8}/L_{\mathrm{TIR}}$, which increases for a harder starlight spectrum.

        \cite{Draine2011a} reported that $\nu L_{\nu,8}/L_{\mathrm{TIR}}$ increases by a factor of 1.57 compared to the \cite{Draine2007} model if the assumed spectrum is replaced by a $2\times10^4$ K blackbody (cut off at 13.6 eV). This factor will only slightly boost the predicted $\nu L_{\nu,8}/M_*$ at the small-$\mathrm{age}_{M_*}$ end, implying a subtle effect on the overall trend of the predicted $\mathrm{age}_{M_*}$--$\nu L_{\nu,8}/M_*$ relation. Furthermore, as the  starlight travels through \hii\ regions before heating the PDR, dust in the \hii\ regions will soften the starlight and reduce the difference between the assumed \cite{Mathis1983} spectrum and the one from a central YSC. Thus, the presence of a harder local radiation field might slightly enhance the PAH emission for YSCs and contribute to the observed scatter, but is not expected to be the main driver of the overall trend observed in Figure \ref{fig:l8_agem}.

    \subsection{The Choice of YSC Catalogs}
    \label{ssec:choices_catalog}

        As mentioned in Section \ref{ssec:ysc_catalogs}, the YSC catalogs provided by \cite{Adamo2017} include 12 catalogs for each galaxy, resulting from the combination of two aperture-correction methods, two stellar libraries, and three extinction/attenuation schemes. The fiducial ones used in this work are derived from SED fitting with the Geneva tracks without rotation and the Milky Way extinction curve of \cite{Cardelli1989} based on photometry from average aperture correction. Here we test whether the choice of YSC catalogs changes our results.

        We take the catalogs generated from the Padova-AGB tracks with the \cite{Cardelli1989} extinction curve and average aperture correction to test the choice of the stellar libraries. Based on the IRAC4 counterparts identified visually, we redo all the calculations except for the IRAC4 photometry. The median and corresponding 1$\sigma$ scatter of the ratios between the new values and the fiducial ones are $1.00_{-0.03}^{+0.02}$ and $1.00_{-0.12}^{+0.03}$ for $\log\mathrm{age}_{M_*}$ and $\log(\nu L_{\nu,8}/M_*)$, respectively. The overall distribution of these sources on the $\mathrm{age}_{M_*}$--$\nu L_{\nu,8}/M_*$ plane remains nearly unchanged. The model predictions of the $\mathrm{age}_{M_*}$--$\nu L_{\nu,8}/M_*$ relations with varying $q_{\mathrm{PAH}}$ or $f_{\mathrm{abs}}$ still cover all the observations for continuous star formation, while two more sources with $\mathrm{age}_{M_*}>40$ Myr move beyond the predicted curve of $q_{\mathrm{PAH}}=4.58\%$ for instantaneous star formation. Therefore, the choice of evolutionary tracks in the SED fitting has no significant impact on our results.

        Similarly, replacing the \cite{Cardelli1989} extinction with the starburst attenuation law of \cite{Calzetti2000} results in small changes around the fiducial values in either $\mathrm{age}_{M_*}$ or $\nu L_{\nu,8}/M_*$ for the two extinction assumptions (i.e, that the ionized gas has either the same or higher attenuation than the stars).

        With regard to the aperture-correction method, the average one changes the normalization of the SED of the YSCs but not the shape, thus only $M_*$ is affected, and age and extinction remain mostly unchanged after the correction \citep{Adamo2017}. Conversely, the CI-based correction changes both the normalization and the shape of the SED, so that $M_*$, age, and extinction are all affected. However, these physical properties derived from the CI-based aperture correction show overall agreement with those derived from the average-based one, with some scatter \citep{Adamo2017,Cook2019}. When we use the catalogs with the CI-based aperture correction, a slightly larger scatter around the fiducial values is found for both $\mathrm{age}_{M_*}$ and $\nu L_{\nu,8}/M_*$. The median and the 1$\sigma$ scatter are $1.00_{-0.05}^{+0.02}$ and $1.00_{-0.06}^{+0.11}$ for $\log\mathrm{age}_{M_*}$ and $\log(\nu L_{\nu,8}/M_*)$, respectively. Such scatters do not significantly change the overall distribution of our sources on the $\mathrm{age}_{M_*}$--$\nu L_{\nu,8}/M_*$ plane. Therefore, our conclusions drawn from the fiducial catalogs are not impacted by the choice of YSC catalogs.

\section{Discussion}
\label{sec:discussion}
    \subsection{PAH Emission Excess}
    \label{ssec:pah_excess}

        In the last section, we have demonstrated that both $q_{\mathrm{PAH}}$ and $f_{\mathrm{abs}}$ play similar roles in predicting the PAH emission. In principle, under the frame of instantaneous star formation, the combination of these two parameters can explain all the observations below the model predictions of the $\mathrm{age}_{M_*}$--$\nu L_{\nu,8}/M_*$ relation presented in Figure \ref{fig:vary_dust}, even for the two we labeled as outliers in Section \ref{sec:results}. For example, assuming $q_{\mathrm{PAH}}=2.50\%$, $f_{\mathrm{abs}}$ should be 0.032 and 0.003 to match the observations of NGC 1313-E 640 and NGC 4395-S 622, respectively.

\begin{figure*}[htb]
        \centering
        \includegraphics[width=0.8\textwidth]{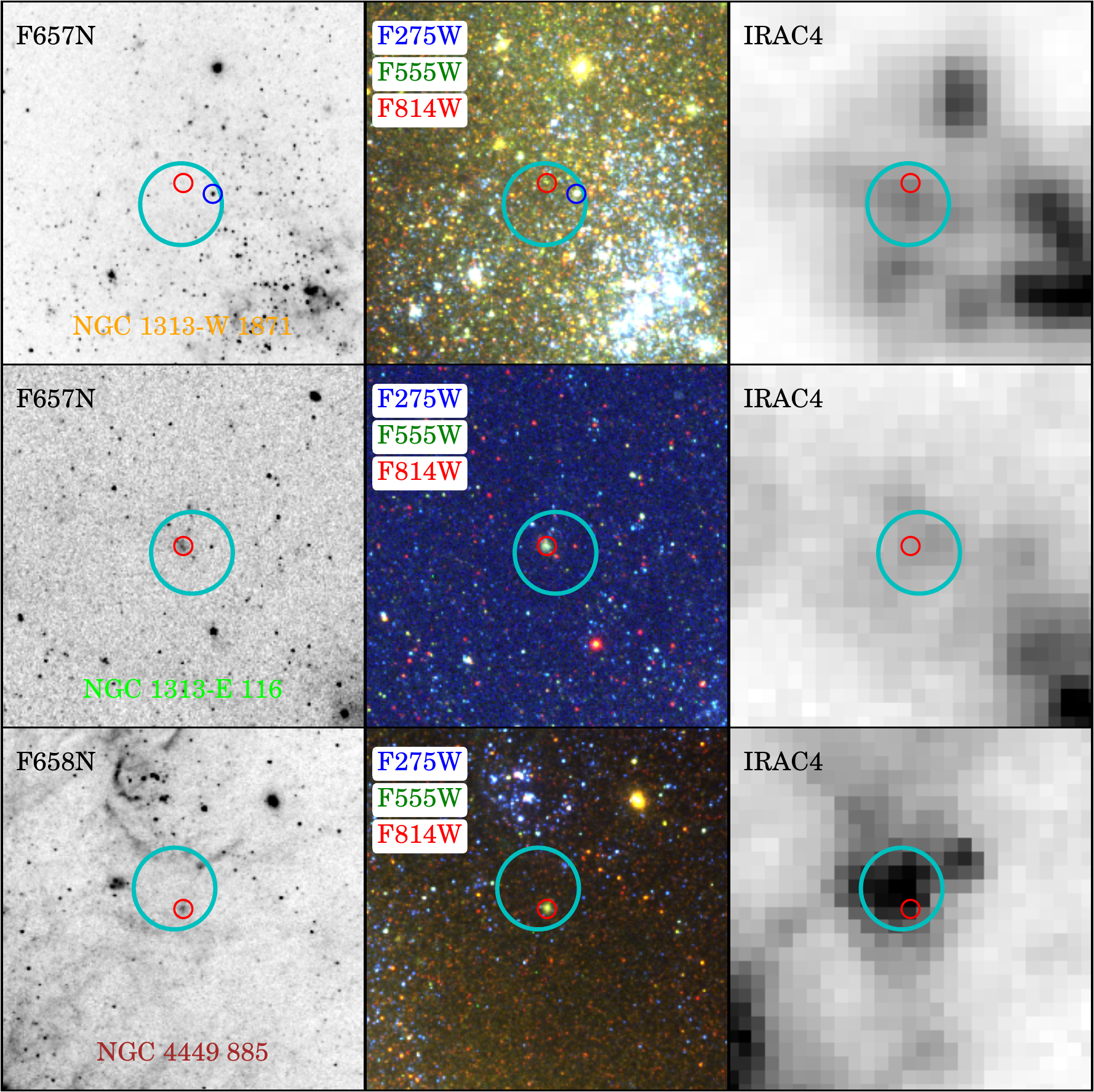}
        \caption{\ha\ (left), optical RGB (middle), and IRAC4 (right) images of three sources with excess PAH emission compared to models. The optical RGB images are created from UV (F275W, blue), $V$ (F555W; green), and $I$ (F814W; red) bands. The symbols are the same as in Figure \ref{fig:outliers} except for the small blue circle in the top row, which is highlighted for discussion. The colors of the source IDs are the same as those of the open circles denoting their locations in the $\nu L_{\nu,8}/M_*$--$\mathrm{age}_{M_*}$ plane in the top panel of Figure \ref{fig:vary_dust}.
        \label{fig:beyond_singleSF}}
        \end{figure*}

        However, for the IRAC4 sources located above the model predictions, it is difficult to reconcile them with our simple model. In Section \ref{ssec:vary_dust} we argue that such PAH excess indicates additional heating sources that might be missed by our YSC selection. In order to understand this PAH excess, we mark three 8 \um\ sources located above the model predictions of instantaneous star formation as colored open circles in the top panel of Figure \ref{fig:vary_dust}. The \ha, optical RGB, and IRAC4 cutouts of these sources are shown in Figure \ref{fig:beyond_singleSF}. Two of them are from NGC 1313, while the other one is from NGC 4449. The \ha\ images are retrieved from the HLA: F657N for NGC 1313 (Program: GO--13773; PI: Rupali Chandar) and F658N for NGC 4449 (Program: GO--10585; PI: Alessandra Aloisi).

        We first attempt to use the identified YSCs alone to explain the PAH emission excess by performing SED fitting with additional \ha\ photometry. There are four clusters within the IRAC4 apertures plotted in Figure \ref{fig:beyond_singleSF}. We measure \ha\ fluxes of these clusters following the photometry method described in \cite{Adamo2017}. Due to the comparable spatial resolution of the \ha\ images to the broadband images of the LEGUS survey, the isolated star cluster samples identified by \cite{Adamo2017} are adopted to construct the growth curves and thus the average aperture corrections. Small offsets between the coordinates of the \ha\ and LEGUS images recovered by the World Coordinate System information are corrected utilizing IRAF tasks. We have applied our photometry code to the LEGUS images and find good agreement between our measurements and those retrieved from the \cite{Adamo2017} catalogs.

        The model photometry for SED fitting is generated from the Yggdrasil population synthesis code \citep{Zackrisson2011} in which a \cite{Kroupa2001} IMF is assumed. To match the catalogs adopted, we use the Geneva tracks without rotation, together with metallicities of $Z = 0.004$, 0.008, and 0.02. The Milky Way extinction curve of \cite{Cardelli1989} is adopted, while $E(B-V)$ varies from 0 to 1.5 in steps of 0.01. For nebular emission, a covering factor of 0.5 is assumed. A $\chi^2$ minimization is implemented to obtain the best-fit model.

        We plot the results from the newly fitted $M_*$ and age, which include \ha\ measurements, in the top panel of Figure \ref{fig:vary_dust} as filled circles with error bars with the same colors as the corresponding open circles. Adding the narrowband fluxes in \ha\ does not significantly change $M_*$ and age for three out of the four clusters. For the first IRAC4 source plotted in Figure \ref{fig:beyond_singleSF} (i.e., NGC 1313-W 1871), two identified YSCs are enclosed by the aperture. Comparing with the best-fit results extracted from the \cite{Adamo2017} catalogs, the central one (NGC 1313-W 1871, marked as red circle) has a similar $M_*$ ($8594_{-1046}^{+1686}~M_{\odot}$ and $9376_{-851}^{+1645}~M_{\odot}$) and age ($200_{-0}^{+100}$ and $400_{-200}^{+0}$~Myr), while the one near the aperture edge (NGC 1313-W 1894, marked as blue circle) shows great changes in both $M_*$ (from $40570_{-29020}^{+3020}~M_{\odot}$ to $6356_{-215}^{+440}$~$M_{\odot}$) and age (from $100_{-85}^{+0}$ to $7_{-0}^{+0}$~Myr). The inclusion of \ha\ fluxes in the SED fitting results in a slightly larger $\mathrm{age}_{M_*}$ but a smaller total $M_*$ compared to the ones based on the \cite{Adamo2017} catalog. According to the new fitting results, NGC 1313-W 1894 ($\mathrm{age_{best}}=7$ Myr) is much younger than NGC 1313-W 1871 ($\mathrm{age_{best}}=400$ Myr). Thus, the 8 \um\ luminosity of that source should be dominated by NGC 1313-W 1894. If we attribute all 8 \um\ emission to NGC 1313-W 1894, the new location of this source is marked by a blue filled circle and an error bar in the top panel of Figure \ref{fig:vary_dust}, which is consistent with the model prediction within the uncertainty. Therefore, with the aid of additional narrowband photometry, the PAH excess of NGC 1313-W 1871 can be explained by a much younger YSC within the same IRAC4 aperture.

        However, for the other two sources (NGC 1313-E 116 and NGC 4449 885), the new fitting results do not significantly alter the original locations. Even if we use the photometry applying the CI-based aperture correction, such PAH excess still exists. NGC 1313-E 116 has an $f_{U,\mathrm{sel}}/f_{U,\mathrm{ap}}$ value of 10\%, which just passes the second criterion listed in Section \ref{sec:results}. If the CI-based photometry is used, $f_{U,\mathrm{sel}}/f_{U,\mathrm{ap}}$ is only 7.6\%, indicating that a large fraction of $U$-band flux is contributed by other stellar objects within the same IRAC4 aperture, and this source would be removed from our final sample. For NGC 4449 885, the CI-based photometry results in a larger age and a smaller $\nu L_{\nu,8}/M_*$ value, but it still lies above the model predictions significantly. Therefore, the identified YSCs alone cannot explain the observed PAH excesses. The ages of the two star clusters covered by these two IRAC4 sources are both $\sim$200--300 Myr and our measurements should not be as high in $\nu L_{\nu,8}$ as observed. We speculate that additional younger (and, possibly, dust-embedded) clusters missed by our YSC selections or detections in the \cite{Adamo2017} catalogs might account for the PAH excess.

        \begin{figure}[htb]
        \centering
        \includegraphics[width=0.48\textwidth]{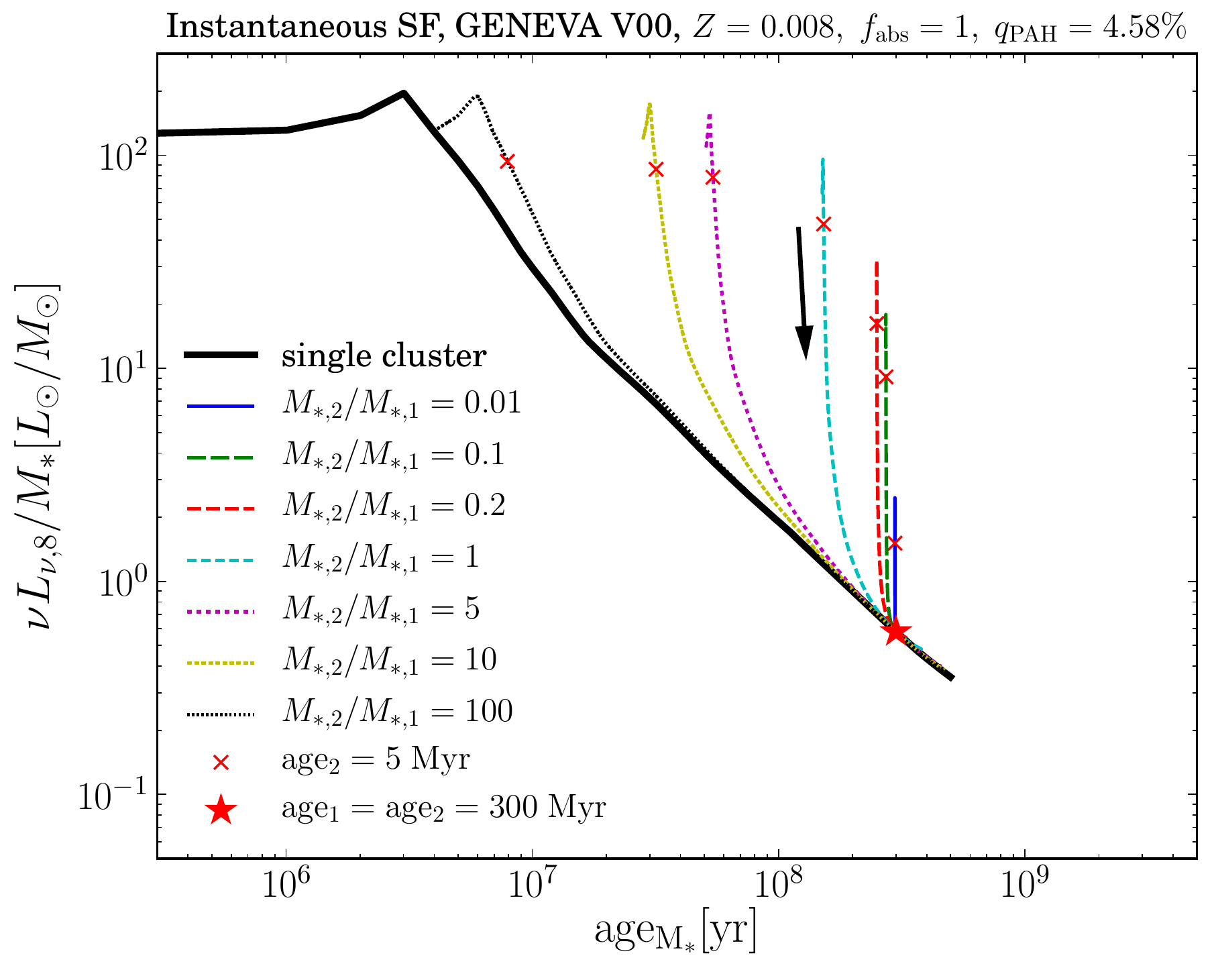}
        \caption{Composite model predictions of the $\mathrm{age}_{M_*}$--$\nu L_{\nu,8}/M_*$ relation for two YSCs with different $M_*$ and ages. The prediction for a single cluster derived from instantaneous star formation (thick black solid curve) is used to calculate $\nu L_{\nu,8}$ for clusters with a given $M_*$ and age. The adopted model parameters are the same as in the top panel of Figure \ref{fig:vary_dust} except that only the model with $q_{\mathrm{PAH}}=4.58\%$ is used. For the two YSCs, the first one (with a subscript of ``1'') is fixed to be 300 Myr old, while the age of the second one (with a subscript of ``2'') varies from 1 to 500 Myr. The combined $\mathrm{age}_{M_*}$ and $\nu L_{\nu,8}/M_*$ are computed for a mass ratio of $M_{*,2}/M_{*,1}=0.01\sim100$. The red ``x'' and star indicate the predictions where the second cluster is 5 and 300 Myr old, respectively, while the black arrow denotes the direction of increasing $\mathrm{age}_2$.
        \label{fig:model_multi_clusters}}
        \end{figure}

        To demonstrate this possibility, we calculate the model predictions of the $\mathrm{age}_{M_*}$--$\nu L_{\nu,8}/M_*$ relation for the combination of two YSCs based on the relation of an instantaneous starburst model with $q_{\mathrm{PAH}}=4.58\%$ (Figure \ref{fig:vary_dust}). The age of the first cluster is set to 300 Myr, which is the same as the two clusters with PAH excess, while the second one has an age that varies from 1 to 500 Myr. The mass ratio of the new clusters changes from 0.01 to 100. In Figure \ref{fig:model_multi_clusters}, we show how the resulting relations vary with different mass ratios. Clearly, the combination of a 300 Myr old cluster with a younger cluster (i.e., the predictions before the red star) will lead to a significant excess of PAH emission, irrespective of the mass ratio. For instance, in the case of $M_{*,2}/M_{*,1}=0.01$, young individual stars/stellar associations with an age of 5 Myr (red ``x'' in Figure \ref{fig:model_multi_clusters}) and $M_*\approx60~M_{\odot}$ and 190 $M_{\odot}$ are massive enough to boost the PAH emission beyond the model significantly for NGC 1313-E 116 and NGC 4449 885, respectively. Several blue point-like sources can be identified within the IRAC4 apertures of NGC 1313-E 116 and NGC 4449 885 by inspection of the optical RGB images in Figure \ref{fig:beyond_singleSF}. Therefore, we expect that such blue stellar objects or others with heavy obscuration might contribute to the observed PAH emission excesses of these two sources.

        Such an excess of PAH emission due to nearby young stellar objects might also happen to YSCs. However, due to the similar age and $M_*/L_I$ to the main clusters, the effect would be naturally included by the mass correction. If the missed stellar objects are older than those used in age-dating, the mass correction becomes problematic and thus the PAH emission excess arises. As a result, the PAH emission excess should be more serious toward high-$\mathrm{age}_{M_*}$, which is what we observe in Figure \ref{fig:vary_dust}.

    \subsection{Missing 8 \um\ Flux}
    \label{ssec:missing_fluxes}

        Our 8 \um\ photometry is performed within the aperture with a radius of 2\farcs25, which corresponds to a physical radius of $\sim$30--50 pc for our selected galaxies. For less massive star associations or clusters, we believe that this aperture is large enough to enclose most of the 8 \um\ fluxes after applying the aperture correction (e.g., \citealt{Hunt2009,Lawton2010}). However, for those super star clusters that are powerful enough to ionize a large \hii\ region, the fixed $\sim$ 50 pc aperture might miss the 8 \um\ flux at some level, leading to an underestimation of $\nu L_{\nu,8}$. For instance, in the most active \hii\ region in the Large Magellanic Cloud (30 Doradus), the distance between PDR clouds and the main ionizing cluster ranges from $\sim$11 to 80 pc \citep{Chevance2016}. Given the current data, the fraction of 8 \um\ flux missed by the fixed aperture is very difficult to estimate and completely degenerate with $f_{\mathrm{abs}}$. The effect of missing the 8 \um\ flux is equivalent to that of varying $f_{\mathrm{abs}}$ and might be another possible reason for the observed scatter of the $\mathrm{age}_{M_*}$--$\nu L_{\nu,8}/M_*$ relation.

    \subsection{Effects of the Undetected Stellar Objects}
    \label{ssec:field_star}

        The mass correction described in Section \ref{ssec:mass_correction} assumes that stellar objects (star clusters and field stars) within the same IRAC4 aperture are same-age stellar populations. If the ages of the undetected stellar objects (e.g., individual field stars or undetected star clusters) are significantly different from the detected ones, the mass correction might be problematic.

        Here, we consider two cases: (a) the detected star clusters are much younger than undetected stellar objects; (b) the detected star clusters are much older than undetected stellar objects. In the first case, only the detected YSCs contribute to the PAH emission, while the total $M_*$ related to the PAH emission will be overestimated after mass correction due to the existence of field stars. Thus, the undetected old stellar objects might be another reason for the observed scatter of the $\mathrm{age}_{M_*}$--$\nu L_{\nu,8}/M_*$ relation. The two outliers shown in Figure \ref{fig:outliers} might belong to this case. In the second case, the observed PAH emission excited by the undetected objects might be attributed to the old star clusters, leading to a mismatch of the excited sources of PAH emission and the derived stellar properties. In fact, this should be the reason for the observed PAH emission excess discussed in Section \ref{ssec:pah_excess}, but only affects those sources with large $\mathrm{age}_{M_*}$. In Appendix \ref{appen:new_fit} we discuss extensively the effects of the undetected stellar populations via the SED fitting using the total fluxes within the aperture.

    \subsection{Upper Limits for Age--PAH Emission Relation}
    \label{ssec:upper_limits}

        We have demonstrated that the $\mathrm{age}_{M_*}$--$\nu L_{\nu,8}/M_*$ relation strongly depends on the PAH abundance and the dust absorption fraction. Even after fixing the PAH abundance using metallicity constraints, it is very difficult to predict the amount of PAH emission for clusters at a given age. It is, however, noteworthy that the model predictions with the highest PAH abundance set an excellent upper limit for the observations, especially for star clusters younger than 10 Myr. Therefore, we fit the model relations derived from a combination of the Geneva tracks with zero rotation and $Z = 0.008$, together with $q_{\mathrm{PAH}}=4.58\%$ and $f_{\mathrm{abs}}=1$, to provide an upper limit to the PAH emission at a given age. Setting $x=\log(\mathrm{age}_{M_*}[\mathrm{yr}])$ and $y=\log(\nu L_{\nu,8}/M_*[L_{\odot}/M_{\odot}])$, we fit the curves with a piecewise function comprising one constant and two second-order polynomials using the Python version of MPFIT\footnote{\url{https://code.google.com/archive/p/astrolibpy/}} \citep{Markwardt2009}. The best-fit results for $\mathrm{age}_{M_*}$ in the range 4--200 Myr are
        
        \begin{equation}\label{eq:best_fit_inst}
            y =
            \begin{cases}
            2.0678, 
            & \hspace{-60pt} 4 < x < 5.6853,\\
            9.9586 - 2.8227x + 0.2524x^2,\\
            & \hspace{-60pt} 5.6853 < x < 6.4786, \\
            19.8870 - 3.8614x + 0.1761x^2, \\
            & \hspace{-60pt} 6.4786 < x < 8.3010
            \end{cases}
        \end{equation}
        and
        \begin{equation}\label{eq:best_fit_cont}
            y =
            \begin{cases}
            2.0960,
            & \hspace{-60pt} 4 < x < 6,\\
            -35.4200 + 11.9823x - 0.9549x^2,\\
            & \hspace{-60pt} 6 < x < 6.6839, \\
            5.3867 - 0.2282x - 0.0415x^2, \\
            & \hspace{-60pt} 6.6839 < x < 8.3010
            \end{cases}
        \end{equation}
        for the instantaneous and continuous star formation models, respectively. These two functions recover the model predictions within 1\%.

        \begin{figure}[htb]
        \centering
        \includegraphics[width=0.48\textwidth]{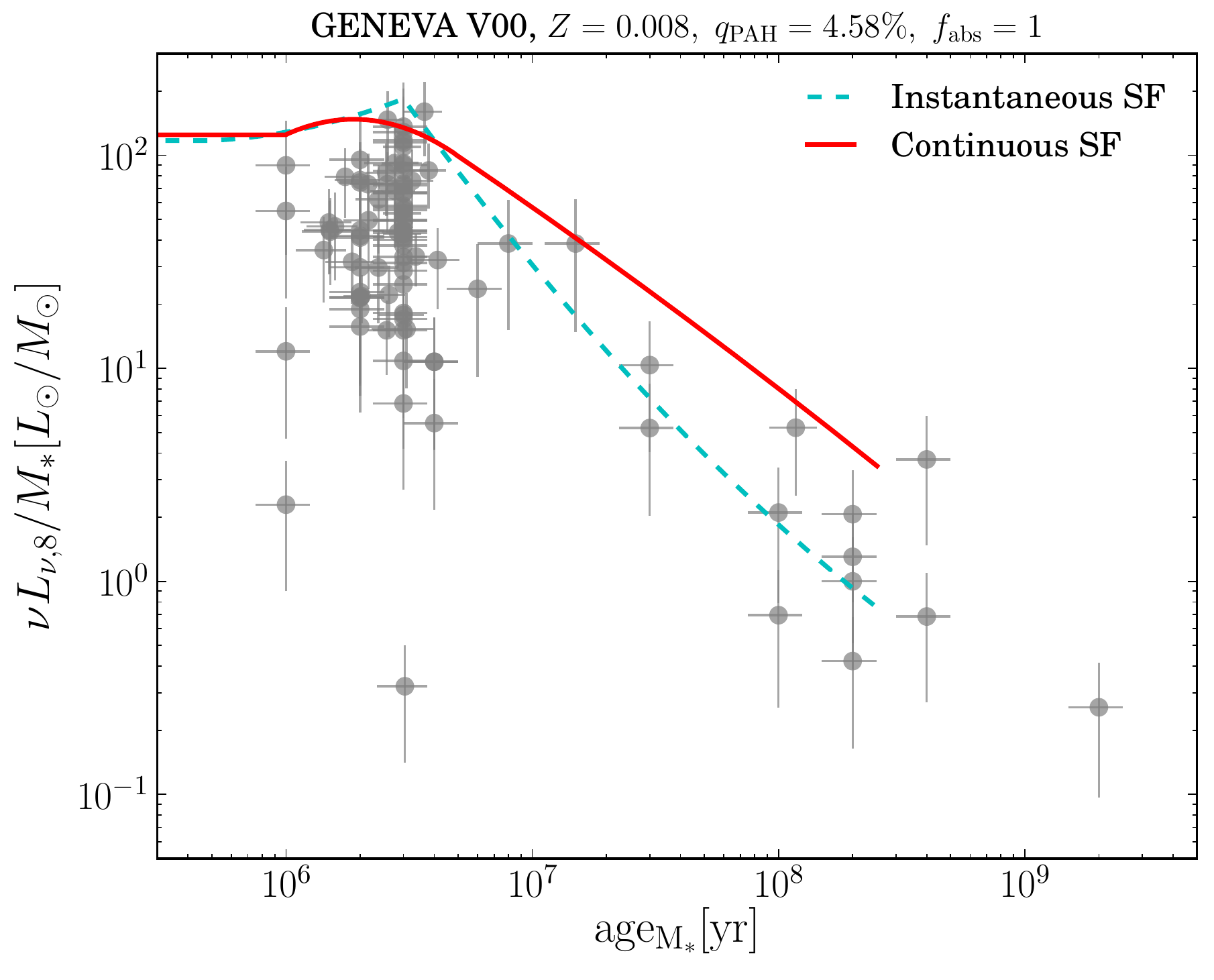}
        \caption{Best-fit results of the predicted $\mathrm{age}_{M_*}$--$\nu L_{\nu,8}/M_*$ relation for the instantaneous and continuous star formation models. The predicted relations are derived from a combination of the Geneva tracks with zero rotation and $Z=0.008$ (IZ53), together with $q_{\mathrm{PAH}}=4.58\%$ and $f_{\mathrm{abs}}=1$. The underlying gray circles with error bars are the same as the data shown in the right panel of Figure \ref{fig:l8_agem}.
        \label{fig:best-fit}}
        \end{figure}

        In Figure \ref{fig:best-fit} we show these best-fit results, overplotted with our sample. The derived function for the continuous star formation model agrees well with the upper limit of our data, while the one from the instantaneous star formation model also gives a good description for the upper boundary of the data. Thus, Equation (\ref{eq:best_fit_inst}) and (\ref{eq:best_fit_cont}) can be used to estimate the maximum of PAH emission for either a single star cluster or a star forming region with either constant or instantaneous SFR at a given age. We stress that these relations are independent of stellar mass and SFR, provided that the measured $\nu L_{\nu,8}$ is normalized by the underlying stellar mass.

\section{Summary}
\label{sec:summary}

    In this work, we construct a sample of YSCs with Spitzer IRAC4 (8 \um) counterparts from five galaxies within 5 Mpc based on the public YSC catalogs from LEGUS \citep{Adamo2017}. We investigate the relation between the mass-weighted ages ($\mathrm{age}_{M_*}$) of star clusters and the mass-normalized 8 \um\ luminosities ($\nu L_{\nu,8}/M_*$), and construct simple models to explain the observed trend and the scatter.

    We obtain an anticorrelation with a Pearson correlation coefficient of $r=-0.84\pm0.05$ between $\mathrm{age}_{M_*}$ and $\nu L_{\nu,8}/M_*$ in which the normalized 8 \um\ luminosities decrease for older star clusters, indicating that dust (and PAHs) heating by older stellar populations result in lower luminosities compared to younger populations. In modeling the trend, we find that the assumed SSP model does not have a significant effect on the strength of the PAH emission due to the nearly unchanged total UV--NIR luminosity for different choices of the evolutionary tracks or stellar metallicities. When continuous star formation with constant SFR is considered, we find a significant enhancement of the PAH emission at an $\mathrm{age}_{M_*}\gtrsim 5$ Myr, below which the strength of PAH emission is similar to that of the instantaneous star formation model.

    The PAH abundance in the dust model and the assumed dust absorption fraction play the same role in determining the $\nu L_{\nu,8}/M_*$; the changes in these two parameters can result in variations of $\nu L_{\nu,8}/M_*$ of about one order of magnitude within the parameter ranges explored in this work (i.e., $q_{\mathrm{PAH}}=0.47\%-4.58\%$ and $f_{\mathrm{abs}}=0.1-1$). A varying $q_{\mathrm{PAH}}$ or $f_{\mathrm{abs}}$ with a continuous star formation model is able to cover nearly all the observations, irrespective of the assumed SFR. The instantaneous starburst model cannot explain the PAH emission excess of several sources older than $10^8$ yr, even if the highest $q_{\mathrm{PAH}}$ and $f_{\mathrm{abs}}$ are assumed.

    Finally, we fit the model predictions for both the instantaneous and continuous star formation scenarios to give analytical descriptions of the upper limits to the data and their trends. The reported formulas are suitable to estimate the maximum PAH emission at a given age.

    The main limitation to this study is the low spatial resolution of the Spitzer IRAC4 images, which prevents the measurement of the PAH emission that arises from the PDR of individual star clusters. Moreover, our selected 8 \um\ sources are biased toward bright sources, especially for old stellar populations, preventing us from obtaining observations much below the upper limits given in Section \ref{ssec:upper_limits} and studying the dependences of the PAH--age relation on other physical properties, such as the gas-phase metallicity (e.g., \citealt{Draine2007a,Smith2007,Remy-Ruyer2015}), ionization level (e.g., \citealt{Madden2006,Gordon2008,Lebouteiller2011}), and luminosity of the \hii\ region \citep{Binder2018}. 

    The JWST will be launched in the near future. The MIRI instrument will be able to observe galaxies at unprecedented sensitivity levels and subarcsecond spatial resolution in the wavelength range of 5--28.5 \um\ \citep{Rieke2015}. First, the degeneracy between $q_{\mathrm{PAH}}$ and $f_{\mathrm{abs}}$ can be disentangled if either $q_{\mathrm{PAH}}$ is determined using MIR spectra observed via MIRI or $f_{\mathrm{abs}}$ is derived from MIR photometry based on the empirical estimation of the total IR luminosity (e.g., \citealt{Boquien2010,Elbaz2010,Lin2016}). Furthermore, a combination of these high spatial resolution observations with spectra of \hii\ regions (e.g., \citealt{Berg2015,Croxall2016,LinZ2017,Hu2018}) or integral field spectroscopy observations \citep{Sanchez2012} in nearby galaxies will enable us to investigate whether and how the PAH--age relation depends on the local environment at the star cluster scale.

\acknowledgments
    The authors would like to thank B. T. Draine for helpful discussions. This work is supported by the National Key R\&D Program of China (2017YFA0402600) and the National Natural Science Foundation of China (NSFC, Nos. 11421303, 11433005, and 11973038). Z. Lin gratefully acknowledges support from the China Scholarship Council (No. 201806340211).

%



\software{APLpy \citep{Robitaille2012}, Astropy \citep{AstropyCollaboration2013,AstropyCollaboration2018}, IPython \citep{Perez2007}, IRAF \citep{Tody1986,Tody1993}, Matplotlib \citep{Hunter2007}, MPFIT \citep{Markwardt2009}, Numpy \citep{Oliphant2006}, Photutils \citep{bradley2019}, Starburst99 \citep{Leitherer1999,Vazquez2005}}



\appendix

\section{CI test for IRAC4 photometry}
\label{appen:ci_test}

    Our 8 \um\ photometry adopts the average aperture correction based on isolated reference samples (see Section \ref{ssec:irac4_photometry}). The reference sample of each galaxy should have an enough number of isolated sources to provide a robust estimate of the correction. However, given the poor spatial resolution of IRAC4 images, the number of 8 \um\ sources that are isolated enough to construct the growth curve and thus the average aperture correction is very limited for each galaxy. Thus, one worry is whether or not the reference sample we select is representative. Here we present a simple test to make sure that the reference samples and our final sample have similar CI distributions.

    The LEGUS pipeline defines CI as the difference between magnitudes measured within a radius of 1 and 3 pixels, i.e., $\mathrm{CI\equiv mag(1~px) - mag(3~px)}$, for star clusters \citep{Adamo2017,Cook2019}. Given that the FWHM of the IRAC4 PSF is $\sim2\arcsec$ ($\sim3$ pixels), such definition might not be suitable for the IRAC4 images. The optical images have a PSF of $\sim 2$ pixels, thus we adopt the optical CI definition in units of the PSF to obtain the scaled IRAC4 CI, which is defined as $\mathrm{CI\equiv mag(1.5~px) - mag(4.5~px)}$.

    \begin{figure}[htb]
    \centering
    \includegraphics[width=0.6\textwidth]{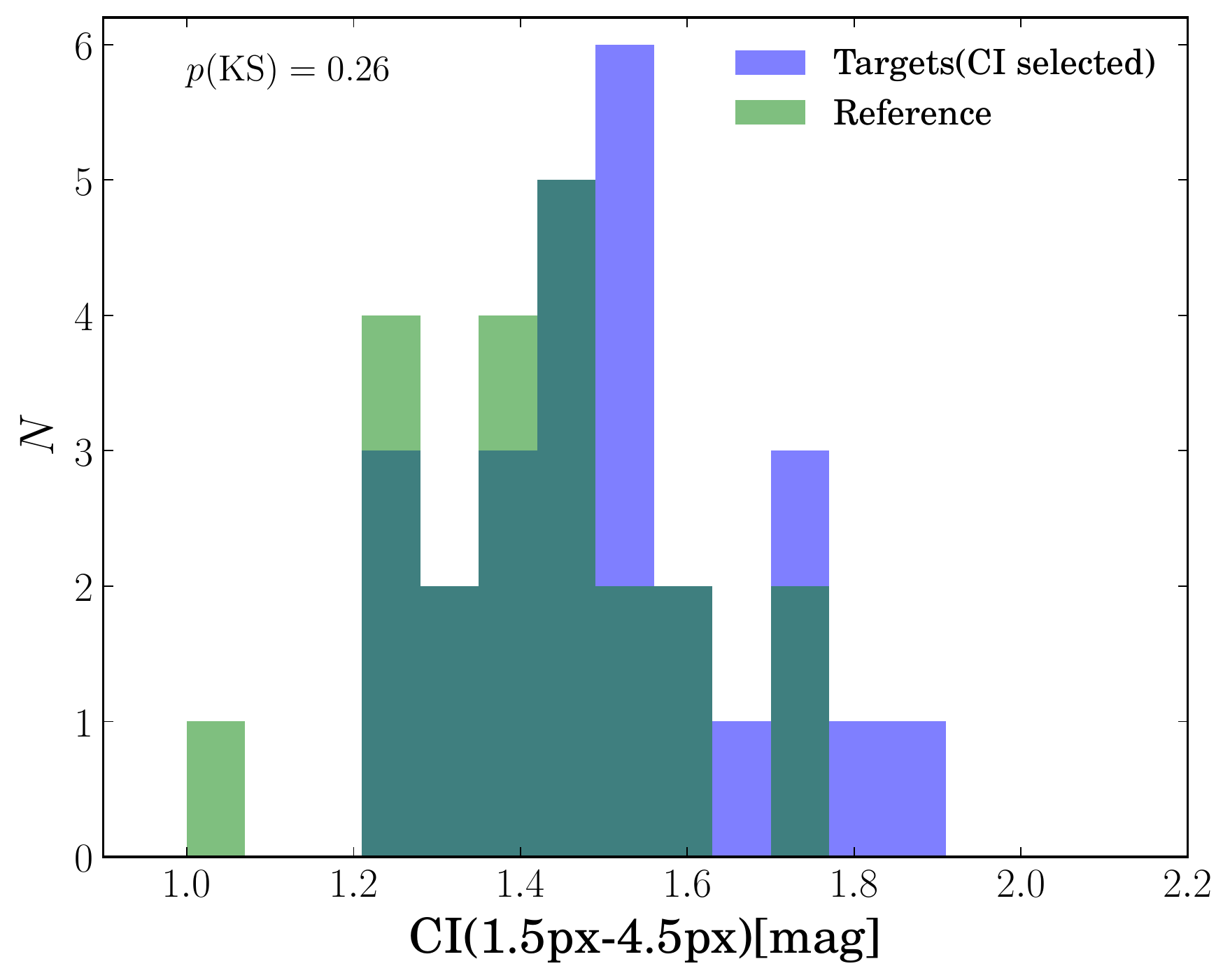}
    \caption{CI distributions for our sample and the reference sample used to construct the growth curves.
    \label{fig:ci}}
    \end{figure}

    In Figure \ref{fig:ci}, we show the CI distributions for our targets and the reference sample. Our targets plotted in Figure \ref{fig:ci} are a subsample of 146 sources within which all sources have a clear profile out to 4.5 pixels to ensure a robust measurement of CI. Although the distributions of the reference sample and the selected targets show slight difference, a Kolmogorov--Smirnov test results in a $p$-value of 0.26 ($>0.05$), suggesting that the difference between the two CI distributions is not significant. Therefore, we believe that the average corrections constructed from this reference sample are representative for our final sample.

\section{SED Fitting with Total fluxes within the IRAC4 Aperture}
\label{appen:new_fit}

    As mentioned in Section \ref{ssec:vary_sfh}, the adopted IRAC4 apertures always contain more than one stellar object, and thus our observations might be better described by complex stellar populations. In Section \ref{ssec:vary_dust} and Section \ref{ssec:vary_fabs}, we show that with varying $q_{\mathrm{PAH}}$ and/or $f_{\mathrm{abs}}$, the continuous star formation models are consistent with almost all the observations. However, all these observations are the combination of stellar properties (i.e, stellar mass and age of star clusters) derived for individual clusters and 8 \um\ luminosities integrated across an aperture of 2\farcs25. Although additional mass corrections (Section \ref{ssec:mass_correction}) and criteria (Section \ref{sec:results}) are applied to reduce the uncertainties arising from the difference in aperture, here we go further to check how the undetected stellar objects within the same aperture influence the estimation of stellar properties and which SFH is a better description of the photometry by performing the SED fitting using the total fluxes within the 2\farcs25 aperture.

    We measure the total fluxes within the same aperture as the one used for the IRAC4 photometry in the images of the five LEGUS broad bands and the \ha\ narrow band. All \ha\ images are retrieved from the HLA: F657N for NGC 7793, NGC 4395, and NGC 1313 (Program GO--13773; PI: Rupali Chandar); F658N for NGC 4449 (Program: GO--10585; PI: Alessandra Aloisi), and NGC 3738 (Program: 9892; PI: Rolf Jansen). These aperture fluxes are directly fed to the SED fitting based on the $\chi^2$ minimization approach. We consider both instantaneous and continuous star formation models for which most of the assumptions are the same as those described in Section \ref{ssec:pah_excess}, except that a fixed stellar metallicity of $Z=0.02$ is adopted. All 146 IRAC4 isolated regions are included in this new measurement and fitting.

    \begin{figure}[htb]
    \centering
    \includegraphics[width=0.8\textwidth]{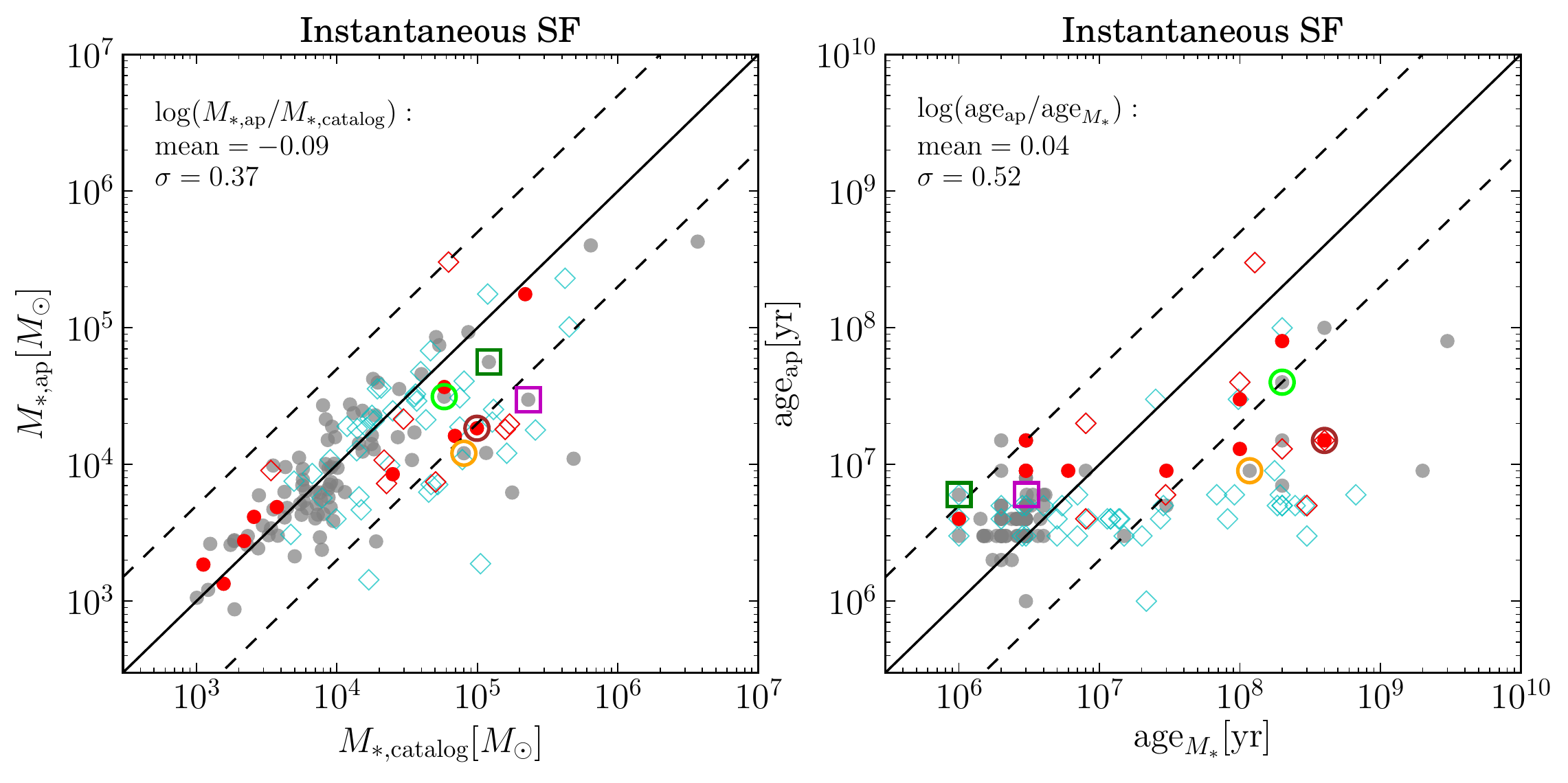}
    \includegraphics[width=0.8\textwidth]{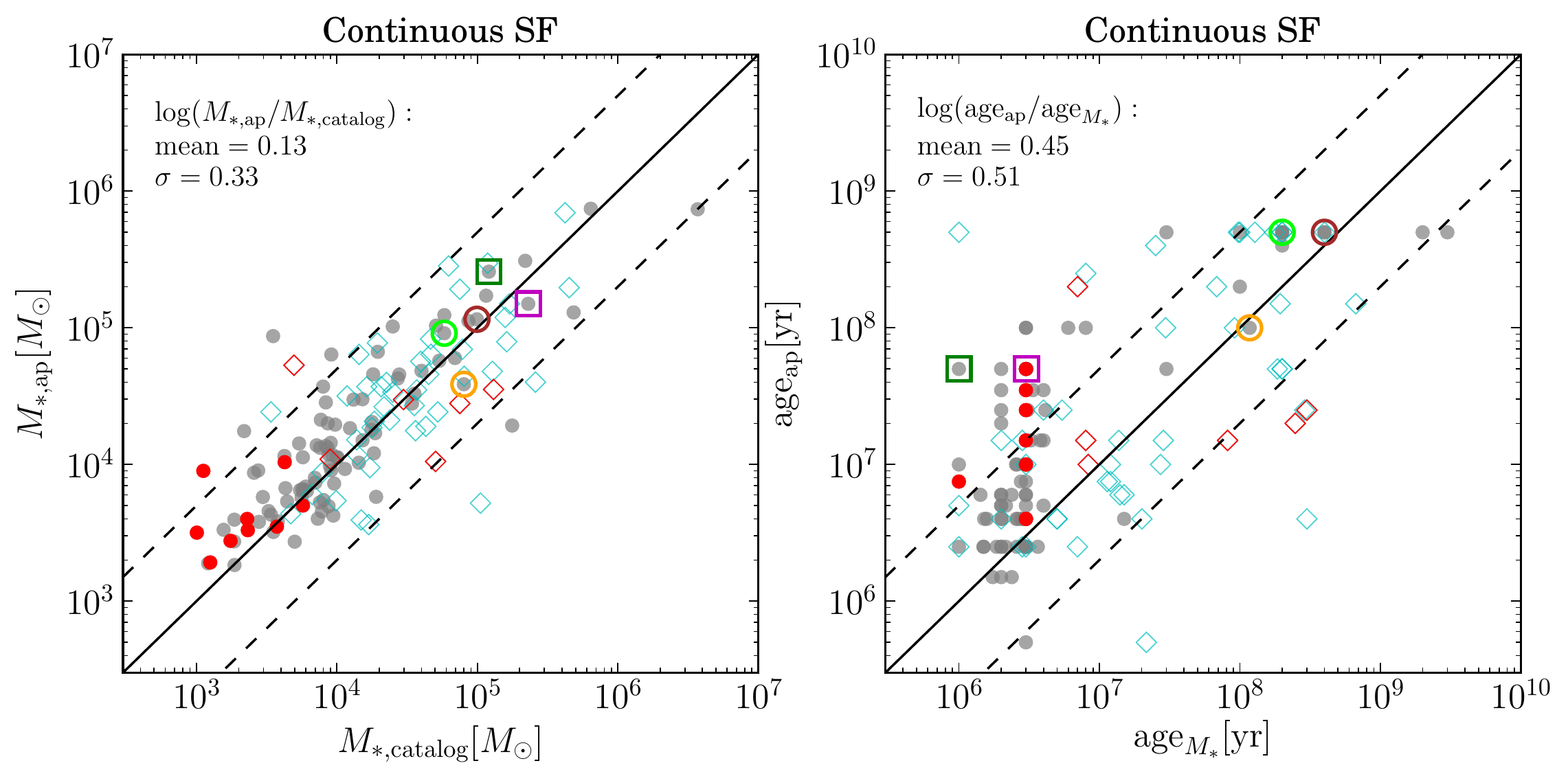}
    \caption{Comparisons of $M_*$ (left) and age (right) between results retrieved from the \cite{Adamo2017} catalogs ($M_{*,\mathrm{catalog}}$ and $\mathrm{age}_{M_*}$) and the new fitting results ($M_{*,\mathrm{ap}}$ and $\mathrm{age_{ap}}$) assuming instantaneous (top) and continuous (bottom) star formation models. The filled circles represent our final sample of 97 sources, while the open cyan diamonds are those sources excluded by three criteria given in Section \ref{sec:results}. The red filled circles and open diamonds indicate sources with acceptable best-fit results (i.e., $Q>0.001$). In each panel, the mean and standard deviation of the ratio between the two parameters for the final sample are listed at the upper-left corner. The open squares are the two outliers shown in Figure \ref{fig:outliers}, while the open circles indicate the three sources located above the instantaneous star formation models shown in Figure \ref{fig:vary_dust} and \ref{fig:beyond_singleSF}. The colors of these symbols are the same as those of the source IDs in the corresponding figures. The black solid lines are the one-to-one relations, and the dashed ones denote a factor of 5 around the one-to-one lines .
    \label{fig:comp_m_age}}
    \end{figure}

    Because of the small uncertainties in our photometry ($\sim$0.03~mag), which are about half the uncertainty in the photometry of individual star clusters ($\sim$0.06~mag), neither the instantaneous nor the continuous star formation model provides acceptable best-fit results (i.e., $Q>0.001$) for most of our sources. Among the 97 sources in our final sample, only 10 and 9 of them have acceptable best-fit results within the uncertainties of the measured aperture photometry for the instantaneous and continuous star formation models, respectively; these numbers are 18 and 15, respectively, for the whole 146 sample. Assuming that our small photometric uncertainties reflect the actual measurements, the high fraction of unacceptable best-fit results might imply that both the instantaneous and continuous star formation models are too simple to describe the stellar populations within the IRAC4 aperture.

    However, it is still valuable to compare the stellar parameters derived from the best fits of individual star clusters with those of our IRAC4 regions. In Figure \ref{fig:comp_m_age}, we present the comparisons of the stellar mass and age between the results based on the \cite{Adamo2017} catalogs ($\mathrm{age}_{M_*}$ and $M_{*,\mathrm{catalog}}$) and those derived from the new SED fitting using the total fluxes within the IRAC4 aperture ($M_{*,\mathrm{ap}}$ and $\mathrm{age_{ap}}$). Sources with acceptable and unacceptable best-fit results are distinguished by different colors. The means and standard deviations of the ratios between the new fitting values and the catalog values for our final sample are also listed in the figure.

    There appear to be no significant differences between sources with acceptable and unacceptable best-fit results in terms of the comparisons between $M_{*,\mathrm{ap}}$ ($\mathrm{age_{ap}}$) and $M_{*,\mathrm{catalog}}$ ($\mathrm{age}_{M_*}$). The additional criteria given in Section \ref{sec:results} significantly reduce the number of mismatched sources that might have large differences in the estimated stellar properties between our defined ones ($\mathrm{age}_{M_*}$ and $M_{*,\mathrm{catalog}}$) and those from the new fitting. For our final sample of 97 sources, the derived $M_{*,\mathrm{catalog}}$ is in good agreement with $M_{*,\mathrm{ap}}$ regardless of the adopted SFHs. Therefore, the mass correction described in Section \ref{ssec:mass_correction} is able to provide reliable mass estimation within the IRAC4 aperture.

    The situation is more complex for the age. When instantaneous star formation is assumed, $\mathrm{age}_{M_*}$ gives a nearly unbiased estimation of $\mathrm{age_{ap}}$ with a slightly large scatter. However, $\mathrm{age}_{M_*}$ tends to overestimate $\mathrm{age_{ap}}$ for sources with large $\mathrm{age}_{M_*}$, including the three sources with the PAH emission excess discussed in Section \ref{ssec:pah_excess}. When continuous star formation models are adopted, our $\mathrm{age}_{M_*}$ tends to underestimate $\mathrm{age_{ap}}$ for several ``young'' sources.

    The measured aperture fluxes are from both the detected star clusters/associations listed in the \cite{Adamo2017} catalogs and the undetected stellar objects (field stars or faint stars clusters). Thus, the differences between $\mathrm{age}_{M_*}$ and $\mathrm{age_{ap}}$ should result from the differences between the ages of the detected and undetected populations. We consider three cases: (a) the detected and undetected populations have nearly the same ages, (b) the detected star clusters are much younger than the undetected stellar objects, and (c) the detected star clusters are much older than the undetected stellar objects. In the first case, for our 8 \um-selected sources, the $\mathrm{age_{ap}}$ derived from both SFHs should be similar to $\mathrm{age}_{M_*}$, and the stellar populations in the aperture should be very young. This represents most of our sources with $\mathrm{age}_{M_*}\sim1$--10 Myr whose $\mathrm{age_{ap}}$ are consistent with $\mathrm{age}_{M_*}$ regardless of the SFHs. In the second case, the detected clusters should be the exciting sources of the observed 8 \um\ emissions. However, the existence of the undetected old stellar populations might lead to an overestimation of the total $M_*$ related to the PAH emission as the undetected populations have little contribution to the PAH emission but significant contribution to the aperture optical fluxes. When continuous star formation is assumed, sources in this case might have an older $\mathrm{age_{ap}}$ compared to $\mathrm{age}_{M_*}$, accounting for those sources with large differences between $\mathrm{age_{ap}}$ and $\mathrm{age}_{M_*}$ shown in the bottom panel of Figure \ref{fig:comp_m_age} and the two outliers shown in Figure \ref{fig:outliers}. In the third case, the optical fluxes from the detected star clusters and the observed PAH emission are mismatched, resulting in the PAH emission excess discussed in Section \ref{ssec:pah_excess} and a smaller $\mathrm{age_{ap}}$ compared to $\mathrm{age}_{M_*}$ when instantaneous star formation is assumed. This represents sources with $\mathrm{age}_{M_*}>100$ Myr which are better described by the continuous star formation models.

    Overall, both assumptions for SFH fall short of providing a satisfactory description of the observed SEDs in the IRAC4 apertures within the nominal uncertainties for most of our sources, at least in terms of quality of fit. However, the comparisons between stellar parameters from the best-fit models of individual clusters and our integrated ones suggest that (a) our mass corrections provide reliable mass estimates within the IRAC4 aperture, and (b) continuous star formation (or multiple stellar populations) still should be taken into account even within our small regions (i.e., $\sim$30--50 pc).

\section{Hints from Metal-poor Galaxies}
\label{appen:metal-poor}

    Previous works revealed a positive correlation between the PAH emission/abundance and metallicity (e.g., \citealt{Engelbracht2005,Madden2006,Draine2007a,Smith2007,Remy-Ruyer2015}). As elucidated in Section \ref{ssec:vary_fabs}, we attribute the observed scatter in the $\mathrm{age}_{M_*}$--$\nu L_{\nu,8}/M_*$ relation to the variations in $f_{\mathrm{abs}}$ due to the narrow range of metallicity of our sample (four out of five galaxies have gas-phase metallicity of $12+\log(\mathrm{O/H})\sim8.4$). Here we present a preliminary attempt of unveiling the role of PAH abundance, which is related to metallicity \citep{Draine2007a,Remy-Ruyer2015}.

    To this end, we relax the galaxy criteria listed in Section \ref{ssec:galaxy_selection} and select galaxies with $12+\log(\mathrm{O/H})\lesssim8.0$ from the LEGUS sample based on the metallicities given in \cite{Calzetti2015a}. Ten galaxies (include NGC 3738) are selected, covering a metallicity range of $12+\log(\mathrm{O/H})=7.82-8.10$ and a distance range of 3.05--6.4 Mpc. YSC selections, 8 \um\ counterpart identification, and image processing are the same as those described in Section \ref{ssec:ysc_catalogs} and Section \ref{sec:image}. Because the number of isolated 8 \um\ clumps is not large enough to construct a reasonable growth curve for these selected galaxies, we adopt the growth curve constructed from NGC 1313 (green dashed--dotted curve in Figure \ref{fig:growth_curves}) for aperture correction. Recall that the choice of growth curve only changes the total 8 \um\ fluxes by up to 20\%. Fifty-five IRAC4 sources are identified and 28 of them remain after applying three criteria given in Section \ref{sec:results}. Hereafter, these 28 sources are denoted as the metal-poor sample ($12+\log(\mathrm{O/H})\sim8.0$), while the 95 sources (after removing 2 sources from NGC 3738) discussed in Section \ref{sec:results} are labeled as the metal-rich sample ($12+\log(\mathrm{O/H})\sim8.4$).

    \begin{figure}[htb]
    \centering
    \includegraphics[width=0.485\textwidth]{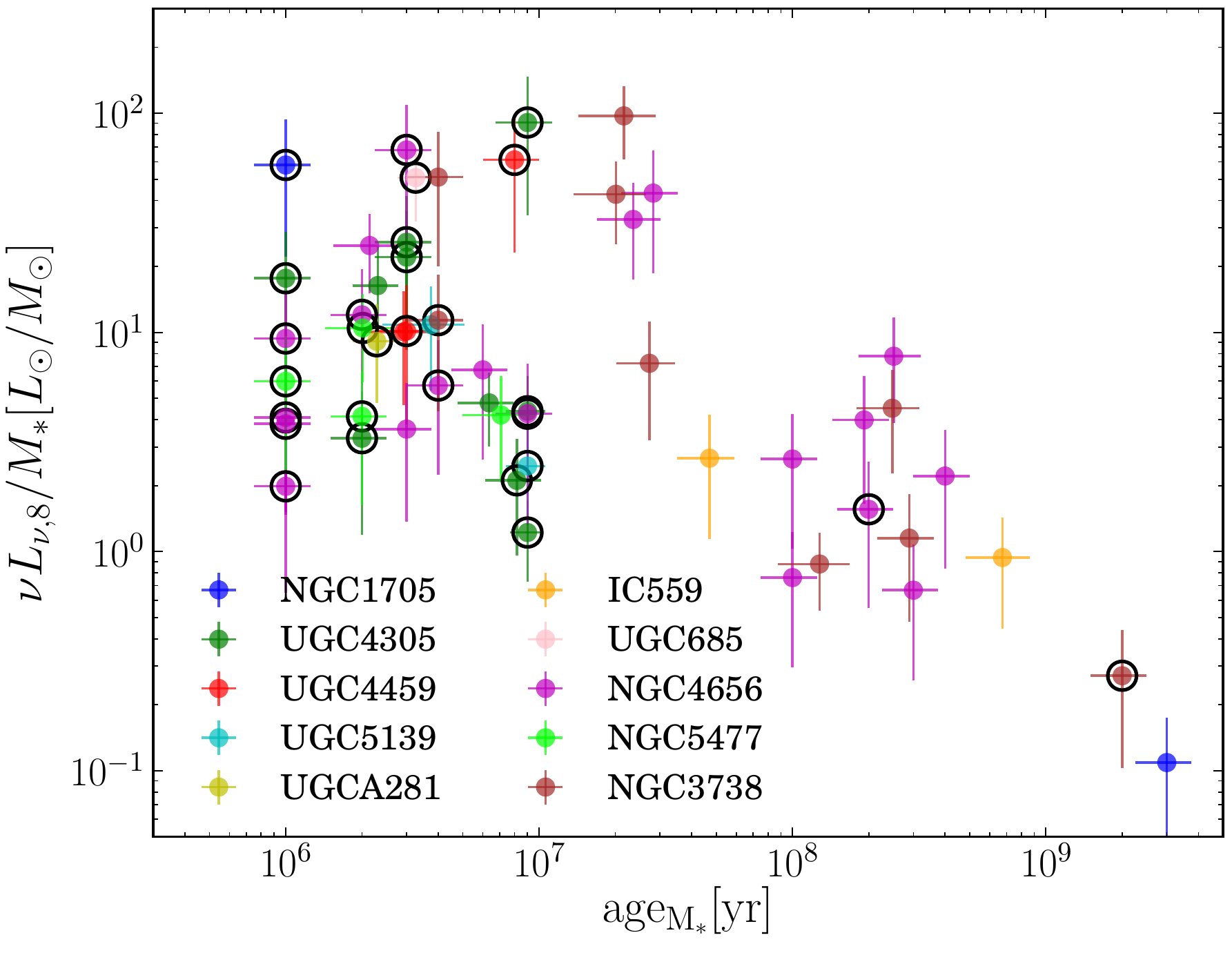}
    \includegraphics[width=0.485\textwidth]{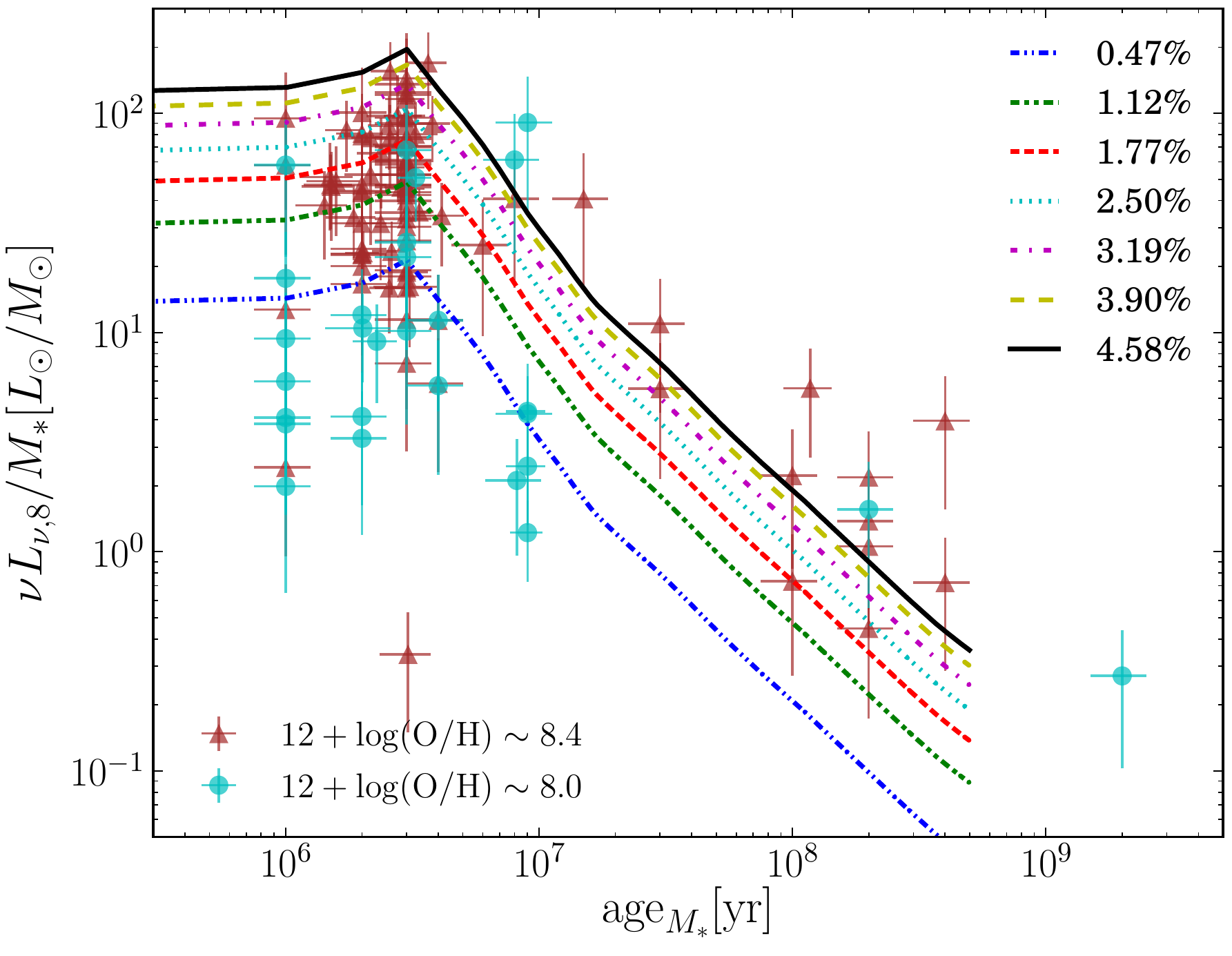}
    \caption{Left: $\nu L_{\nu,8}/M_*$ as a function of $\mathrm{age}_{M_*}$ for all 55 IRAC4 sources identified from 10 metal-poor galaxies. The black open circles indicate 28 sources in the metal-poor sample. Right: comparison between the metal-rich sample (brown triangles) and the metal-poor sample (cyan circles). Model predictions from different dust models shown in the top panel of Figure \ref{fig:vary_dust} are also given.
    \label{fig:comp_metalpoor}}
    \end{figure}

    In the left panel of Figure \ref{fig:comp_metalpoor}, we plot the $\mathrm{age}_{M_*}$--$\nu L_{\nu,8}/M_*$ relation for all 55 sources from 10 metal-poor galaxies and mark out the final metal-poor sample with black open circles. Similar to the metal-rich population shown in Figure \ref{fig:l8_agem}, the whole metal-poor population shows an anticorrelation between $\nu L_{\nu,8}/M_*$ and $\mathrm{age}_{M_*}$ although the scatter is still large. We compare the distribution of the metal-poor sample in the $\mathrm{age}_{M_*}$--$\nu L_{\nu,8}/M_*$ plane to the metal-rich sample in the right panel of Figure \ref{fig:comp_metalpoor}. For $\mathrm{age}_{M_*}\lesssim 10$ Myr, the metal-poor sample tends to have smaller $\nu L_{\nu,8}/M_*$ at fixed $\mathrm{age}_{M_*}$ compared to the metal-rich sample. For $\mathrm{age}_{M_*}>10$ Myr, the size of the metal-poor sample is too small to draw any reliable conclusion.

    \begin{figure}[htb]
    \centering
    \includegraphics[width=0.7\textwidth]{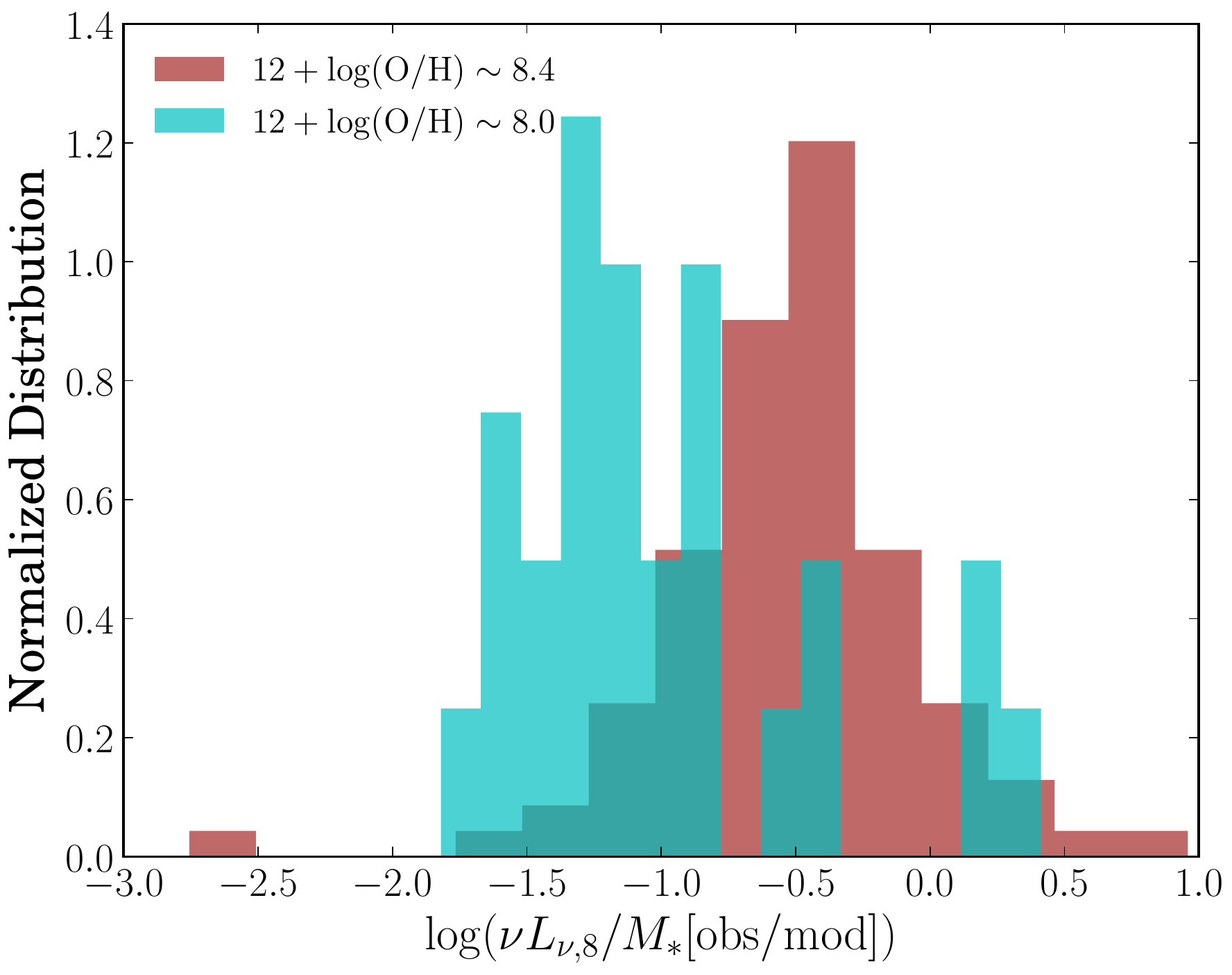}
    \caption{Normalized distributions of the ratios between the observed $\nu L_{\nu,8}/M_*$ and the predicted values from the model with $q_{\mathrm{PAH}}=4.58\%$ for the metal-rich sample (brown) and metal-poor sample (cyan).
    \label{fig:hist_metalpoor}}
    \end{figure}

    In order to display the difference clearly, we take the predicted $\mathrm{age}_{M_*}$--$\nu L_{\nu,8}/M_*$ relation derived from the instantaneous star formation model and $q_{\mathrm{PAH}}=4.58\%$ as the fiducial model, and calculate the ratios between the observed $\nu L_{\nu,8}/M_*$ and the predicted values given the observed $\mathrm{age}_{M_*}$ ($\nu L_{\nu,8}/M_*$[obs/mod]). The normalized distributions of $\nu L_{\nu,8}/M_*$[obs/mod] for two samples are shown in Figure \ref{fig:hist_metalpoor}. Obviously, the metal-poor sample has weaker PAH emission compared to the metal-rich sample, while the medians and the 68\% scatters of $\nu L_{\nu,8}/M_*$[obs/mod] are $0.07_{-0.04}^{+0.27}$ and $0.32_{-0.19}^{+0.38}$, respectively. However, one should note that besides $q_{\mathrm{PAH}}$, $f_{\mathrm{abs}}$ is another important parameter that drives the variation in the PAH emission. Both observations \citep{Boquien2009,Reddy2010,Qin2019} and semianalytical model \citep{Cousin2019} suggest more dust absorption for more metal-rich galaxies. Therefore, the observed reduction of the PAH emission for the metal-poor sample might be attributed to the decrease in either $q_{\mathrm{PAH}}$ or $f_{\mathrm{abs}}$. Given the current data, we cannot determine which parameter is predominant.




\bibliography{ms_PAH}






\end{document}